\documentclass[12pt,preprint]{aastex}

\usepackage{amsmath}
\newcommand{\xp}[2]{\frac{\partial #1}{\partial #2}}

\newcommand{\tabfrac}[2]{%
        \setlength{\fboxrule}{0pt}%
        \fbox{$\frac{#1}{#2}$}%
}

\shorttitle{Second order spin-orbit resonances}

\shortauthors{Flynn \& Saha}

\begin{document}

\title{Second order perturbation theory for spin-orbit resonances}
\author{Angela E. Flynn and Prasenjit Saha}
\affil{Astronomy Unit, School of Mathematical Sciences, \\
Queen Mary, University of London, Mile End Road, London E1 4NS, UK}

\email{a.e.flynn@qmul.ac.uk}
\email{p.saha@qmul.ac.uk}

\begin{abstract}

We implement Lie transform perturbation theory to second order for the
planar spin-orbit problem. The perturbation parameter is the
asphericity of the body, with the orbital eccentricity entering as an
additional parameter. We study first and second order resonances for
different values of these parameters. For nearly spherical bodies
like Mercury and the Moon first order perturbation theory is adequate,
whereas for highly aspherical bodies like Hyperion the spin is mostly
chaotic and perturbation theory is of limited use. However, in
between, we identify a parameter range where second order perturbation
theory is useful and where as yet unidentified objects may be in
second order resonances.

\end{abstract}


\section{Introduction} \label{Introduction}

The discovery by \citet{pettengillanddyce} of the $3:2$ spin-orbit
resonance of Mercury ignited interest in rotational-orbital dynamics.
Mercury remained until recently the only solar-system body in an
asynchronous spin-orbit resonance. \citet{colombo} proffered an
explanation for Mercury's unusual fate; this work was soon followed by
the seminal papers of \citet{goldreichandpeale1,goldreichandpeale2} and
much later those of \citet{celletti-part1,celletti-part2}, all of whom
undertook a detailed study of spin-orbit dynamics.

Renewed attention was afforded to the subject with the article by
\citet{wpm} on the nature of Hyperion's rotation. These authors asserted
that the irregularly shaped satellite tumbles chaotically as it revolves
around Saturn. Unambiguously confirmed shortly thereafter by ground-based
observations \citep{klavetter}, Hyperion's chaotic state is often
invoked to illustrate a prime example of chaotic dynamics in
action. Many other satellites have seen episodes of chaotic rotation in
their past. \citet{wisdom1987} showed that any object that occupies the
$1:1$ state must have crossed a chaotic zone at some point in its
history. If the zone is attitude-stable as is generally the case for
uniformly shaped bodies, trapping is temporary and the body escapes to a
regular spin state. Otherwise the fate is inevitably one of long-term
non-principal axis rotation.

In addition to the primary resonances, the phase space of a nonlinear
dynamical system may contain secondary, or even more complicated
resonances. Motivated by the tidal heating conundrum concerning
Enceladus, \citet{wisdom2004} developed a theoretical foundation of
secondary configurations. He proposed that capture into the $3:1$
secondary resonance could provide a plausible mechanism for the
resurfacing activity on Enceladus and its apparent role as a source of
E-ring material.

Further, as the name suggests, instances of spin-orbit coupling may have
significant effects on the orbital motion of the bodies
concerned. \citet{blitzer1979} was among those who -- recognizing the
mathematical parallels between mean motion and spin-orbit
commensurabilities -- proposed that a unified theory be developed to
encompass both forms of interaction. Like their mean motion
counterparts, spin-orbit resonances result in a stabilizing effect on
the motion of objects bound in their domain. Moreover, examination of
the whole resonant structure provides insight into the past evolution of
the system. Additionally, the nature of the coupling can shed light on
the future dynamics. \citet{chauvineauandmetris1994} address this oft
forgotten issue showing that in the case of an ellipsoidal satellite
orbiting its parent planet at a distance of approximately 10 times the
satellite's mean radius, such resonances can lead to chaotic orbital
motion.

Though we possess a good understanding of capture into orbital
resonances, the mechanics behind capture into a spin-orbit resonance
remains relatively unknown. \citet{cellettietal1998} write that ``no
mathematical proofs are nowadays available on the effective mechanism of
capture into a resonance, but it is widely accepted that one cannot
explain the actual state of the satellites without invoking dissipative
torques.'' Thus while the inconclusiveness of current methods is freely
acknowledged, most authors tend to favor those approaches adopted by
both Darwin and Macdonald as described in \citet{goldreichandpeale1}. Indeed, no alternative mechanism had been proposed until recently, when the work of \citet{correiaandlaskarnature} unraveled the outstanding question concerning Mercury's capture into the $3:2$ state.

\citet{correiaandlaskarnature} argue that for any eccentricity, the spin
rate of a body is naturally driven towards some equilibrium value which
depends on its current eccentricity. Since the eccentricity varies due
to orbital perturbations, the spin rate can be naturally pumped both up
and down. Crossing a resonance has some probability of resulting in the 
capture of the object and preventing further evolution of the spin. An untrapped 
spin rate can cross a resonance repeatedly, thus increasing its probability
of being trapped. However, a captured body can be released if the
eccentricity falls low enough for that resonance to become unstable to
tidal torque.

The sort of complex interaction between orbit and spin seen in recent
work make it interesting to study other resonances. In particular,
\citet{correiaandlaskarnature} point out that Mercury's present
equilibrium spin rate is close to the $5:4$ resonance, even though the
actual spin rate is trapped in the $3:2$ state. But the $5:4$ is a
second order resonance and does not appear in a simple first order
treatment of spin-orbit dynamics, even though it is easy to find in
numerical integrations. Accordingly we are motivated in this paper to
develop a perturbative treatment of the planar spin-orbit problem to
second order. We use the now-standard technique of Lie
transforms. Through Lie transform perturbation theory we can derive an
integral of motion for both non-resonant and resonant orbits, and use
this integral to generate composite surfaces of section. These we can
easily compare with numerical surfaces of section. The algebraic
details quickly become messy, especially at second order, and we will
not show them in full in this paper\footnote{Full details, including programs and computer-algebra scripts, are available from the authors on request.}. Instead, we will illustrate the procedure in detail for a driven double pendulum, which is closely analogous to a spin-orbit system but much simpler.

This paper is organized as follows: in
Sect.~\ref{TheoryforSpin-OrbitCoupling} we present the Hamiltonian
formulation of the spin-orbit problem. Also given is the Hamiltonian
of the driven pendulum whose dynamics provide an analog with the
spin-orbit coupling. Sect.~\ref{PerturbationTheory} sees development
of the perturbative approach through Lie transforms, including
application to resonances of first and second order. In
Sect.~\ref{Results} we investigate how our perturbative model fares in
relation to numerical integrations for selected solar system
bodies. Sect.~\ref{Conclusions} comprises a brief summary of the paper.

\section{Theory for Spin-Orbit Coupling}
\label{TheoryforSpin-OrbitCoupling}

\subsection{The Spin-Orbit Hamiltonian}
\label{TheSpin-OrbitHamiltonian}

\begin{figure}[h]
\epsscale{0.8}\plotone{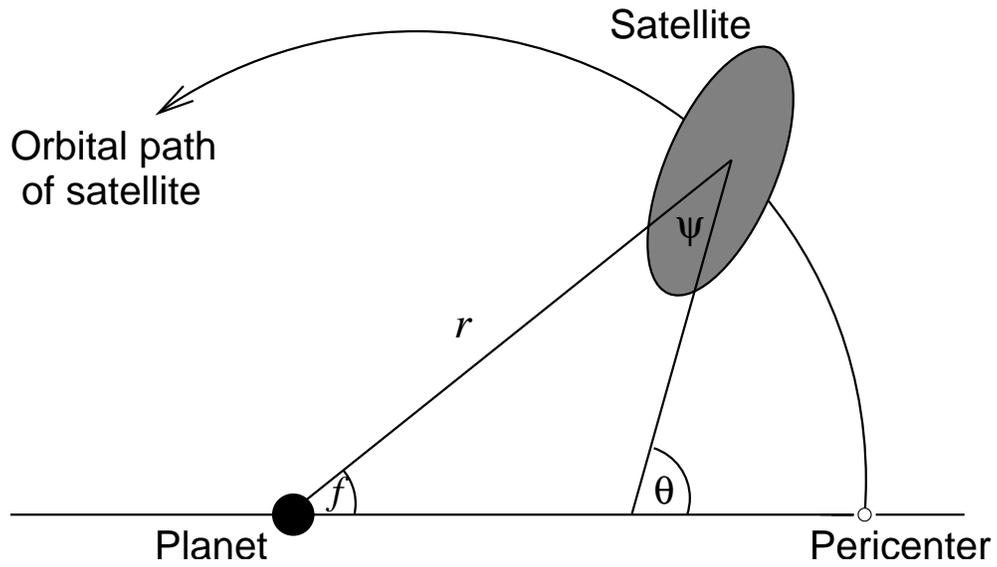}
\caption{Geometry of a body in a spin-orbit resonance: $\theta$ is the
angle between the ellipsoid's longest axis and a reference line,
chosen for the sake of simplicity to coincide with the semi-major axis
of the satellite's fixed orbit.
\label{fig1}}
\end{figure}

The spin-orbit problem is discussed in modern texts such as
\citet{murrayanddermott} or \citet{sussmanandwisdom}, so here we only
cover the essentials briefly. In the planar problem, pictured in
Fig.~\ref{fig1}, the angle $f$ is the true anomaly while $\psi$ is the
angle between the satellite-planet line and central axis of the
satellite. It is clear that $\theta=f+\psi$. Neglecting tidal torques,
the resulting equation of motion for $\theta$ is
\citep{goldreichandpeale1}
\begin{equation}
\label{fullpendulum_eq} 
\ddot{\theta} =
-{\frac{1}{2}}\,\alpha^2\,n^2\,{\frac{a}{r^{3}}}\,\sin{(2\,(\theta-f))}
\end{equation}
where $a$ is the satellite's semi-major axis, $r$ measures the
satellite's radial distance to the planet, and $n$ is the mean motion
(i.e. angular orbital frequency). The asphericity parameter $\alpha$
is defined in terms of the moments of inertia $A\le B \le C$ by
\begin{equation} 
\label{alpha} 
\alpha = \sqrt{\frac{3\,(B-A)}{C}}.
\end{equation}
The equation of motion Eq.~(\ref{fullpendulum_eq}) is explicitly
integrable in two cases: (i)~when the satellite is oblate i.e. $A=B$,
there is zero torque indicating a freely rotating body, or (ii)~when
the orbit is circular, which corresponds to a pendulum equation.

It is easy to see that Eq.~(\ref{fullpendulum_eq}) is equivalent to
the Hamiltonian
\begin{equation}
\label{newhamiltonian}
H(p,\,q,\,t) = \frac{p^2}{2} - \frac{\alpha^2}{4\,r^{3}}\,\cos{(2\,(q-f))}
\end{equation}
where $f=f(t)$, $r=r(t)$, and $q\equiv\theta$. Units have been chosen
such that $a=n=1$ and hence $2\,\pi$ corresponds to an orbital
period. The Hamiltonian can be rendered autonomous by the standard
device of extending the phase space. The equivalent autonomous
Hamiltonian is
\begin{equation}
\label{revisedhamiltonian}
H(p_1,\,p_2;\,q_1,\,q_2) = \frac{p_{1}^2}{2} + p_2
- \frac{\alpha^2}{4\,r^{3}}\,\cos{(2\,(q_1-f))}.
\end{equation}
Here $q_1=\theta, p_1=\dot\theta, q_2=t$ (and also equals the mean
anomaly in celestial mechanics), while $p_2$ has a value such that
$H(p_1,\,p_2;\,q_1,\,q_2)=0$.

The Hamiltonian (Eq.~(\ref{revisedhamiltonian})) depends on $q_2$
through $r$ and $f$. For perturbation theory we wish to make the
$q_2$-dependence explicit. There are classical techniques
for doing so, but here we introduce a new way which is well-suited to
computer algebra.

Consider the expansion
\begin{equation}
\label{fourier}
\frac{1}{r^3}\,\exp{(2\,f\,i)} = \sum_{k}\,H_{k}(e)\,\exp{(k\,{q_2}\,i)},
\end{equation}
which is a Fourier series in $q_2$ with Fourier coefficients
\begin{equation}
\label{fouriercoeff}
H_{k}(e) = \frac{1}{2\,\pi}\,\int_{0}^{2\,\pi}{\frac{1}{r^3}\,
\exp{((2\,f-k\,{q_2})\,i})}\,{d{q_2}}.
\end{equation}
Let us recall the standard expressions for Keplerian motion in terms
of the auxiliary variable $E$ (called the eccentric anomaly in
celestial mechanics):
\begin{eqnarray}
\label{rtf}
r   &=& 1-e\,\cos{E}, \nonumber \\
q_2 &=& E-e\,\sin{E}, \\
\tan{\frac{f}{2}} &=& \sqrt{\frac{1+e}{1-e}}\,\tan{\frac{E}{2}}. \nonumber
\end{eqnarray}
From Eq.~(\ref{rtf}) we have $d{q_2}=r\,dE$, which lets us rewrite
Eq.~(\ref{fouriercoeff}) as
\begin{equation}
\label{Hde}
H_{k}(e) = \frac{1}{2\,\pi}\,\int_{0}^{2\,\pi}{\frac{1}{r^2}\,
\exp{((2\,f-k\,{q_2})\,i})}\,{dE},
\end{equation}
thus making the integrand an explicit function of the integration
variable. The integral (Eq.~(\ref{Hde})) may be solved as a power series
in $e$ for any given $k$; the integration is somewhat tedious by hand,
but trivial using computer algebra. The answer will always be
real. Now, multiplying the complex conjugate of Eq.~(\ref{fourier}) by
$\exp{(2\,{q_1}\,i)}$ and taking the real part gives
\begin{equation}
\label{poisson}
\frac{1}{r^3}\,\cos{(2\,(q_{1}-f))} =
\sum_{k}\,H_{k}(e)\,\cos{(2\,{q_1}-k\,{q_2})}.
\end{equation}
The potential part of the Hamiltonian (Eq.~(\ref{revisedhamiltonian}))
is now expressed as a Fourier series in $q_1,q_2$ with coefficients
$H_{k}(e)$ given by Eq.~(\ref{Hde}). The $H_{k}(e)$ themselves are
power series in the eccentricity. They are credited in the literature
to \citet{cayley} and we will refer to them as {\em Cayley
coefficients}. They are tabulated in Table~1 of
\citet{goldreichandpeale1} and Table~2.1 of \citet{celletti-part1}.

Thus the two degree of freedom autonomous Hamiltonian is
\begin{equation}
\label{primaryH}
H(p_1,\,p_2;\,q_1,\,q_2) = \frac{{{p_1}}^2}{2} + {p_2} - 
  \frac{\alpha^2}{4}\,\sum_{k = k_{\rm min}}^{{k_{\rm max}}}\,
       H_{k}(e)\,      \cos{(2\,{q_1} - k\,{q_2})}.
\end{equation}
Under this guise, the $1:1$ resonance with $k=2$ has the argument
$\cos{(2\,{q_1}-2\,{q_2})}$; similarly, $k=3$ for the $3:2$
commensurability, and so the associated argument is
$\cos{(2\,{q_1}-3\,{q_2})}$.

A body with low $e$ generally has as its final spin state the
synchronous lock: this is dictated by the form of the Cayley
coefficient $H_{2}(e)$, which in the limit as $e\rightarrow0$, tends
towards unity. Moreover the leading term in the $e-$series decays as
${\it O}(e^{\vert{k-2}\vert})$. So for small values of the
eccentricity, it is reasonable to focus on the dominant resonances,
namely the $1:1$ and the $3:2$, though several adjacent resonances
must also be considered, not least to permit study of the effects
caused by small divisors. For increasing eccentricity however, the
presence of higher order $e$ terms in the expansion is crucial --
these are now the prominent terms, substantially larger than their
counterparts at lower order in $e$. Indeed, the higher the
eccentricity, the longer it takes the series to converge.

\placetable{tbl1}

To illustrate the preceding argument, in Table~\ref{tbl1} we have
tabulated the Cayley coefficients for selected solar system bodies. We
also include an example of a highly eccentric object. While Mercury's
eccentricity is usually considered to be rather sizable, as far as this
analysis is concerned, it falls into the low $e$ category. On the other
hand, Nereid's exceptionally high eccentricity of $e=0.75$ affords
substantial significance to Cayley coefficients at growing distance from
the traditionally strongest coupling; that is, the synchronous state.

\subsection{Driven Pendulum}
\label{DrivenPendulum}

Our perturbative calculations with the Hamiltonian of Eq.~(\ref{primaryH}) contain a minimum of four terms in the sum. At second order, the resulting expressions span several pages. Rather than mask the underlying simplicity of the perturbative scheme in such a mess, we choose instead a simple example to illustrate the mechanics of perturbation theory in the form of Lie transforms. The driven pendulum is a double pendulum -- see Fig.~\ref{fig2} -- with inner pendulum driven at unit angular frequency. Its Hamiltonian is
\begin{equation}
\label{drivenH}
H(p_1,\,p_2;\,q_1,\,q_2) = \frac{p_{1}^2}{2} + p_2 
-{\alpha}\,\cos{q_1} -{\beta}\,\cos{(q_{1}-q_2)}.
\end{equation}
which is quite similar to the spin-orbit Hamiltonian, with the angles
$q_1$ and $q_2$ being analogous to the orientation angle and orbital
phase (mean anomaly) respectively. For a derivation of
Eq.~(\ref{drivenH}) see the Appendix. A different driven pendulum,
having one more driving term, is considered in
\citet{sussmanandwisdom}.

\section{Perturbation Theory}
\label{PerturbationTheory}

\subsection{Resonant Integrals}
\label{ResonantIntegrals}

An important property of Hamiltonian systems is the following: If $H$
depends on $q_1$ and $q_2$ only through the combination $l\,q_1-k\,q_2$,
then ${k}\,{p_1}+{l}\,{p_2}$ will be a constant of the motion. Various
forms of this statement appear in the literature -- see for instance
Theorem~2 of \citet{gustavson}, but here we provide a simple proof.

Given any coprime integers $k,\,l$, we can always find integers $i,\,j$,
such that the matrix
\begin{eqnarray}
\label{Smatrix}
M=
\begin{pmatrix}
i  & j  \\
k  & l 
\end{pmatrix}
\end{eqnarray}
has unit determinant. We can then define
\begin{eqnarray}
\label{relateP}
\begin{pmatrix}
P_1  \\
P_2 
\end{pmatrix}
&=&
\begin{pmatrix}
i  & j  \\
k  & l 
\end{pmatrix}
\begin{pmatrix}
p_1  \\
p_2 
\end{pmatrix} \\
\label{relateQ}
\begin{pmatrix}
Q_1 \\
Q_2 
\end{pmatrix}
&=&
\begin{pmatrix}
l  & -k  \\
-j & i 
\end{pmatrix}
\begin{pmatrix}
q_1  \\
q_2
\end{pmatrix}.
\end{eqnarray}
The integer matrix of Eq.~(\ref{relateQ}) is the inverse transpose of
that in Eq.~(\ref{relateP}). This transformation is canonical (see
e.g. Sect.~3 of \citet{binneyandspergel}).

Now if $H$ depends on $Q_1={l}\,q_1-{k}\,q_2$ but is cyclic in $Q_2$,
then $P_2={k}\,p_1+{l}\,p_2$ will be a constant of the motion, often
called the {\em fast action}. $Q_1$ is the resonant (and hence
slowly varying) angle and $Q_2$ is fast by comparison. As
$(P_1,\,P_2)$ form the conjugate pair to $(Q_1,\,Q_2)$ they acquire
the same nomenclature; thus $P_1$ is the slow or resonant
action. Following averaging over $Q_2$, the fast action $P_2$ is
constant.

Let us illustrate the conservation of $P_2$ by considering the Hamiltonian 
\begin{equation}
H=\frac{1}{2}\,p_1^2 + p_2 + {\kappa}\,\cos{(n\,(l\,q_1-k\,q_2))}
\end{equation}
with $H=0$. This gives us 
\begin{equation}
\label{fastexample}
p_2 = -\frac{1}{2}\,p_1^2 - {\kappa}\,\cos{(n\,(l\,q_1-k\,q_2))}.
\end{equation}
Making use of this equation and our earlier result that
${k}\,p_1+{l}\,p_2={\rm constant}$ we have
\begin{equation}
\label{tidy}
\frac{1}{2}\,\left(p_1-\frac{k}{l}\right)^{2}+
{\kappa}\,\cos{\left(n\,l\,\left(q_1-\frac{k}{l}\,q_2\right)\right)}
= {\rm constant}
\end{equation} 
which is a pendulum equation.  If we now introduce a resonant angle
$\gamma \equiv q_1-(k/l)\,q_2$, we will recover the pendulum equation in
Eq.~(5) of \citet{goldreichandpeale1}.

The well-known overlap criterion of \citet{chirikov} corresponds to
the overlap of oscillations of pendulum equations resulting from two
or more resonance terms. This is considered a diagnostic for the
onset of large-scale chaos.

\subsection{Lie Transforms}
\label{LieTransforms}

Hamiltonian perturbation theory is based on transforming a Hamiltonian
$H(p,\,q)$ into a so-called Kamiltonian $K(p',\,q')$ which is easier
to solve. Lie transforms are the standard technique for carrying out
the desired transformation, and are described in several books, e.g.
\citet{sussmanandwisdom}. We summarize the Lie transform method
here, following \citet{cary}.

Let us consider a transformation $T$ whose effect is
\begin{equation}
\label{operator}
T\,K(p,\,q) = K(p',\,q')
\end{equation}
where $K$ represents an arbitrary functional form. In particular,
\begin{equation}
T\,{p} = p', \quad T\,{q} = q'.
\end{equation}
Now, if we define 
\begin{eqnarray}
H(p,\,q)=K(p',\,q')
\end{eqnarray}
then $T\,K=H$ essentially changes the functional form of the new
function $K$ to the functional form of the old function, $H$.

Thus far, $T$ could be any transformation. But for perturbation theory
we are interested in a canonical transformation $T$ derived from a
generating function $W$, depending on a perturbation parameter
$\epsilon$ such that as $\epsilon\rightarrow0$, $W\rightarrow0$ and 
$T\rightarrow1$. Accordingly, we now introduce the generating function
\begin{equation}
W=\epsilon\,W_1+{\epsilon^2}\,W_2
\end{equation}
and define linear operators $L_1$ and $L_2$ such that
\begin{equation}
\label{poissonbracket}
L_1\,F \equiv \sum_{i}^{n}\left(\xp{F}{q_{i}}\xp{W_1}{p_{i}}-
                         \xp{F}{p_{i}}\xp{W_1}{q_{i}}\right)
\end{equation}
and similarly for $L_2$. Here $n$ is the number of degrees of freedom.
The transformation $T$ is now expressed as the operator
\begin{equation}
\label{liegeneral}
T = 1+\epsilon\,L_1 + {\epsilon^2}\,\left(\case{1}{2}\,L_1^2 + L_2\right)+ \ldots.
\end{equation}
To ${\it O}(\epsilon^2)$, $K$ is related to $H$ as follows:
\begin{eqnarray}
\label{prelim}
T\left(K_0+{\epsilon}\,K_1 + {\epsilon^2}\,K_2\right) &=& H_0+{\epsilon}\,H_1 + {\epsilon^2}\,H_2, \nonumber \\
\label{KversusH}
\left(1+{\epsilon}\,L_1+{\epsilon^2}\,\left(\tabfrac{1}{2}L_1^2 + L_2\right)\right)\left(K_0+{\epsilon}\,K_1 + {\epsilon^2}\,K_2\right) &=& H_0+{\epsilon}\,H_1 + {\epsilon^2}\,H_2.
\end{eqnarray}
Equating powers of $\epsilon$ we obtain the series of equations
\begin{eqnarray}
\label{K0versusH}
K_0 &=& H_0,\\
\label{K1versusH}
L_1 K_0 &=& H_1 - K_1,\\
\label{K2versusH}
L_2 K_0 &=& H_2 - \tabfrac{1}{2}L_1 \left(H_1 + K_1\right) - K_2.
\end{eqnarray}
In a first order perturbative calculation we choose $K_1$ so that it
contains only secular and resonant terms (and hence has a constant of
motion, which can be calculated) and construct a $W_1$ so as to solve
Eq.~(\ref{K1versusH}). At second order, we choose $K_2$ so as to
contain only secular and resonant terms, and construct a $W_2$ so as to
solve Eq.~(\ref{K2versusH}). The appropriate choice of $K_1$ and $K_2$
avoids small denominators.

At this point it is worth clarifying our use of subscripts. For the most
part, subscripts ${0,\,1,\,2}$ respectively indicate zeroth, first, and
second order terms. On the other hand, the subscripts on the generalized
coordinates $(p_1,\,p_2;\,q_1,\,q_2)$ are unrelated to order: rather
they were introduced upon extension of the phase space.

\subsection{Example: Driven Pendulum}

We now work through an example to illustrate how to implement Lie
transform perturbation theory, illustrating how a perturbative model
of the driven pendulum fares in relation to numerical integration in
Fig.~\ref{fig3}. We find that the perturbative solution approximates
well both the location and extent of the islands.

The Hamiltonian for the driven pendulum was defined in
Eq.~(\ref{drivenH}). For algebraic convenience we make the
replacements
$\alpha\rightarrow\epsilon\,\alpha,\beta\rightarrow\epsilon\,\beta$ (and
set $\epsilon=1$ for numerical work), which lets us write
\begin{eqnarray}
H_0 &=&\frac{p_{1}^2}{2} + p_2, \nonumber \\
H_1 &=&-{\alpha}\,\cos{q_1} -{\beta}\,\cos{(q_{1}-q_2)}, \\
H_2 &=& 0.
\end{eqnarray}

\subsubsection{Driven pendulum: $1:1$ resonance}

Let us start by considering a first order resonance. $H_0$ by
construction is integrable. From Eq.~(\ref{K0versusH})
\begin{equation}
\label{iamh0}
K_0=H_0=\frac{p_{1}^2}{2} + p_2.
\end{equation}
Clearly $K_0$ is cyclic in $q_1$ and $q_2$. $K_1$ is chosen such that
$H_1-K_1$ will have no resonant or secular terms. In other words, we
take any terms in $H_1$ that are either independent of $q_1,\,q_2$ or involve
the resonant argument (in this case $q_1-q_2$) and copy them into $K_1$.
Thus we choose
\begin{equation}
\label{K1_11}
K_1(p_1,\,p_2;\,q_1,\,q_2) =-\beta\,\cos{({q_1}-{q_2})}.
\end{equation}
With this choice Eq.~(\ref{K1versusH}) becomes
\begin{equation}
\label{minusalphacosq1}
{L_1}{K_0}=-{\alpha}\,\cos{q_1}. 
\end{equation}
To solve this equation for the first order generating function $W_1$, we
first observe that if
\begin{equation}
\label{W1format}
W_1 = -\frac{A\,\sin{(m\,q_1+n\,q_2)}}{m\,p_1+n}
\end{equation}
and $K_0$ is given by Eq.~(\ref{iamh0}), then
\begin{equation}
\label{L1K0format}
L_1\,K_0 = A\,\cos{(m\,q_1+n\,q_2)}.
\end{equation}
Comparing Eqs.~(\ref{L1K0format}) and (\ref{minusalphacosq1}), and
then plugging into Eq.~(\ref{W1format}) gives
\begin{equation}
\label{W1to1fullo1}
W_{1}(p_1,\,p_2;\,q_1,\,q_2) = \frac{\alpha\,\sin{({q_1})}}{{p_1}}.
\end{equation}
Note that had we opted to assign $K_1=0$ rather than the value given
in Eq.~(\ref{K1_11}) then $W_1$ would have included another term (see
Eq.~(\ref{W2to1fullo1})) with denominator $p_1-1$. Clearly this would
introduce a singularity in the transformation at exact resonance, and
a small denominator near resonance. This is of course the problem of
small divisors which in this instance we have avoided.

Though the $1:1$ is a first order resonance, we develop the theory to
second order. Following the above choices for $K_1$ and $W_1$ we have
\begin{equation}
\label{ohneK2}
H_2-\case{1}{2}\,L_1(H_1+K_1) = \frac{\alpha^2}{4\,{{p_1}}^2} - 
  \frac{\alpha^2\,\cos{(2\,{q_1})}}
   {4\,{{p_1}}^2} - 
  \frac{\alpha\,\beta\,\cos{(2\,{q_1} - 
       {q_2})}}{2\,{{p_1}}^2} + 
  \frac{\alpha\,\beta\,\cos{({q_2})}}
   {2\,{{p_1}}^2}.
\end{equation}
We choose $K_2$ to consist of all secular and resonant terms in this
expression, following the same principle as we did in choosing $K_1$.
In fact, there is only one secular term involved here, giving us
\begin{equation}
\label{K2}
K_2 = \frac{\alpha^2}{4\,{{p_1}}^2}.
\end{equation}
Having thus chosen $K_2$, the right-hand side of Eq.~(\ref{K2versusH})
will consist of the last three terms in Eq.~(\ref{ohneK2}).
We now solve Eq.~(\ref{K2versusH}) term by term for $W_2$, just as
we solved Eq.~(\ref{K1versusH}) for $W_1$. We get
\begin{equation}
\label{W21to1pre}
W_2 = \frac{\alpha^2\,\sin{(2\,q_1)}} {8\,p_1^3} 
    + \frac{\alpha\,\beta\,\sin{(2\,q_1 - q_2)}}{2\,p_1^2\,(2\,p_1-1)}
    - \frac{\alpha\,\beta\,\sin{(q_2)}}{2\,p_1^2}
\end{equation}
and rewriting quotients as partial fractions gives
\begin{eqnarray}
\label{W21to1}
W_2 &=& \frac{\alpha^2\,\sin{(2\,{q_1})}}
   {8\,{{p_1}}^3} - 
  \frac{\alpha\,\beta\,\sin{(2\,{q_1} - 
       {q_2})}}{{p_1}} + \nonumber \\
  && \frac{2\,\alpha\,\beta\,\sin{(2\,{q_1} - 
       {q_2})}}{2\,{p_1}-1} - 
  \frac{\alpha\,\beta\,\left( \sin{(2\,{q_1} - 
         {q_2})}+ \sin{({q_2})}
       \right) }{2\,{{p_1}}^2}.
\end{eqnarray}
Again, we see the presence of a small denominator in the third term in
$W_2$. However, this particular factor is acceptable since it is
associated with the $2:1$ resonance, and is thereby outside the domain
of the $1:1$ resonance.

Thus far $H(p_1,\,p_2;\,q_1,\,q_2)$ has essentially undergone a
canonical transformation to produce $K(p'_1,\,p'_2,\,q'_1-q'_2)$.  From
Sect.~\ref{ResonantIntegrals} we know that for the $1:1$ resonance we
are now considering, $p'_1+p'_2$ will be the constant fast action.
Since $T\,p=p'$ we may derive an expression for this fast action
\begin{equation}
C(p_1,\,p_2;\,q_1,\,q_2) \equiv T(p_1+p_2) = {\rm constant}.
\end{equation}
The leading order fast action $C_0$ is simply $p_1+p_2$.  Using the
condition $H(p_1,\,p_2;\,q_1,\,q_2)=0$ to eliminate $p_2$ we have
\begin{equation}
\label{zerothaction1to1}
C_{0} = {p_1} - \frac{{{p_1}}^2}{2} + \alpha\,\cos{({q_1})}+
  \beta\,\cos{({q_1} - {q_2})}.
\end{equation}
In averaging theory the $\alpha\,\cos{({q_1})}$ term would be
jettisoned, since it is a non-resonant or fast term. The first order
correction to the fast action is
\begin{equation}
\label{firstaction1to1}
C_{1} = - \frac{\alpha\,\cos{({q_1})}} {{p_1}}.
\end{equation}
Similarly, $C_2$ is the second order contribution. Observe that the
first order terms are proportional to $\alpha$ and $\beta$ whereas at
second order cross-terms are introduced.
\begin{eqnarray}
\label{secondaction1to1}
C_{2} &=& \frac{-\alpha^2\,\left( 2 + 
         \cos{(2\,{q_1})}  \right) }
     {4\,{{p_1}}^3} + 
  \frac{\alpha\,\beta\,\cos (2\,{q_1} - 
       {q_2})}{{p_1}} - 
  \frac{2\,\alpha\,\beta\,\cos (2\,{q_1} - 
       {q_2})}{2\,{p_1}-1} \nonumber \\
&&+ 
  \frac{\alpha\,\beta\,\left( \cos (2\,{q_1} - 
         {q_2}) + \cos ({q_2})
       \right) }{2\,{{p_1}}^2}.
\end{eqnarray}
Thus the complete expression for the fast action at second order is given by
\begin{equation}
\label{completefastaction}
C = C_0 + C_1 + C_2.
\end{equation}

\subsubsection{Driven pendulum: $2:1$ resonance}

The $2:1$ resonance is second order, so called because the perturbation
must be extended to second order in order to produce it. The resonant
argument $2\,q_1-q_2$ is absent from $H_1$, hence we choose
\begin{equation}
\label{K1_11again}
K_1(p_1,\,p_2;\,q_1,\,q_2) = 0
\end{equation}
Now Eq.~(\ref{K1versusH}) becomes
${L_1}{K_0}=H_1-K_1=-{\alpha}\,\cos{q_1}-{\beta}\,\cos{(q_{1}-q_2)}$.
We solve this equation for $W_1$, taking our cue from 
Eqs.~(\ref{W1format}) and (\ref{L1K0format}) as before, obtaining
\begin{equation}
\label{W2to1fullo1}
W_{1}(p_1,\,p_2;\,q_1,\,q_2) = \frac{\alpha\,\sin{({q_1})}}{{p_1}} + 
  \frac{\beta\,\sin{({q_1} - {q_2})}}
   {{p_1}-1}.
\end{equation}
$K_2$ is composed of the appropriate secular and resonant terms again:
\begin{equation}
K_2 =\frac{-\beta\,\left( -\beta + 
         \alpha\,\cos{(2\,{q_1} - {q_2})} \right)}{4\,
     {\left({p_1}-1 \right) }^2} + 
  \frac{\alpha\,\left(\alpha - 
       \beta\,\cos{(2\,{q_1} - {q_2})}
       \right) }{4\,{{p_1}}^2}.
\end{equation}
Solving Eq.~(\ref{K2versusH}) for this case gives
\begin{equation}
\label{W22to1}
W_2 =  \frac{\alpha^2\,\sin{(2\,{q_1})}}
   {8\,{{p_1}}^3} + 
  \frac{\beta^2\,\sin{(2\,{q_1} - 
       2\,{q_2})}}{8\,
     {\left({p_1}-1 \right) }^3} - 
  \frac{\alpha\,\beta\,\sin{({q_2})}}
   {4\,{\left({p_1}-1 \right) }^2} - 
  \frac{\alpha\,\beta\,\sin{({q_2})}}
   {4\,{{p_1}}^2}.
\end{equation}

The $2:1$ resonant variables are obtained in the same manner as for
$1:1$. In this instance, the leading order fast action has the form
\begin{equation}
\label{zerothaction2to1}
C_{0} = p_1 -{p_1^2} + 2\,\alpha\,\cos{(q_1)} + 
     2\,\beta\,\cos{(q_1-q_2)}.
\end{equation}
The first order correction is
\begin{equation}
\label{firstaction2to1}
C_{1} = -{\frac {\alpha\,\cos{\left( {q_1} \right)}}{{p_1}}}+{\frac {\beta\,\cos{
 \left( {q_1-q_2} \right) }}{{p_1}-1}}.
\end{equation}
As before, $C_2$ is the second order contribution,
\begin{eqnarray}
\label{secondaction2to1}
C_{2} &=& \frac{-\left( \alpha^2\,\left( 2 + 
         \cos (2\,{q_1}) \right) \right)}{4\,{{p_1}}^3} + 
  \frac{\beta^2\,\left( 2 + 
       \cos{(2\,{q_1} - 2\,{q_2})}
       \right) }{4\,{\left({p_1}-1   \right) }^3} - 
  \frac{\alpha\,\beta\,\cos{(2\,{q_1} - 
       {q_2})}}{{p_1}-1} + \nonumber \\
&&
  \frac{\alpha\,\beta\,\cos{(2\,{q_1} - {q_2})}}{{p_1}} + 
  \frac{\alpha\,\beta\,\left(\cos{(2\,{q_1} - 
         {q_2})} + \cos{({q_2})}
       \right) }{2\,{\left({p_1}-1\right) }^2} + \nonumber \\
&&  \frac{\alpha\,\beta\,\left( \cos{(2\,{q_1} - 
         {q_2})} + \cos{({q_2})}
       \right) }{2\,{{p_1}}^2}.
\end{eqnarray}

\subsubsection{Driven pendulum: Surface of section}

In Fig.~\ref{fig3} we show a comparison between second order
perturbation theory and a numerical integration of the Hamiltonian
Eq.~(\ref{drivenH}).

The curves in the upper panel are curves of the second order fast
action $C_0+C_1+C_2$ at $q_2$. In the upper part of the plot -- the
vicinity of the $1:1$ resonance -- the individual terms come from
Eqs.~(\ref{zerothaction1to1}) to (\ref{secondaction1to1}). In the
middle part of the plot -- the neighborhood of the $1:2$ resonance -- the
individual terms come from Eqs.~(\ref{zerothaction2to1}) to
(\ref{secondaction2to1}). The lower part of the plot is the $0:1$
`resonance' region; here the terms in $C_0,\,C_1,\,C_2$ are calculated in
the same way, but we omit the details for brevity.

To produce the lower panel we started a numerical integration from a
random point on each of the perturbative curves and then followed an
orbit for several crossings of the plane $q_2=0$, plotting the value
of $(p_1,\,q_1)$ at each crossing.

The analogous figure for first order perturbation theory would contain
the contours of $C_0+C_1$ at $q_2=0$. Finally, the analogous figure
for averaging theory would have contours of $C_0$ with any
non-resonant periodic terms discarded, also at $q_2=0$.

\section{Results}
\label{Results}

We now proceed to the spin-orbit Hamiltonian (Eq.~(\ref{primaryH})),
defining the perturbation parameter as $\epsilon={\alpha^2}/4$. We set
$k_{\rm min}=1$ and $k_{\rm max}=4$, except for one case (Hyperion)
when we increase $k_{\rm max}$ to 6.

\placetable{tbl2}

In Table~\ref{tbl2} we list the spin-orbit parameters for selected
objects. In fact $\alpha$ exceeds unity for a significant fraction of
solar system bodies, including many asteroids; for instance, 4179
Toutatis has $\alpha\approx1.35$ and 243 Ida has $\alpha\approx1.44$.
Bearing in mind that perturbation theory is only valid for small
$\epsilon$, it is important to find what regimes of $\alpha,e$ our
perturbative model is useful for.

\subsection{The useful regime for perturbation theory}

Fig.~\ref{fig4} is a sketch of the different regimes of $\alpha,\,e$.
(The curves in this figure are not precise boundaries; they are
approximate indications based on examining the results of perturbation
theory and numerical integration for many different parameter values.)
The labeled regions are as follows.

\begin{description}
\item [\hbox to 0.32in {A:}] For $\alpha\lesssim0.05$ or $e\lesssim0.1$ phase space consists of non-resonant spins and first order resonant islands, with no significant second order resonances or chaos. Averaging gets the structure qualitatively right, while first and second order perturbation theory make some quantitative improvement.
\item [\hbox to 0.32in {B:}] For $0.05\lesssim\alpha\lesssim0.2$ and $e\gtrsim0.1$ there are both first and second order islands. Second order perturbation theory successfully recovers the second order islands, whereas averaging and first order perturbation theory are limited to the first order islands.
\item [\hbox to 0.32in {C:}] For $0.2\lesssim\alpha\lesssim0.3$ and $e\gtrsim0.1$ there are significant chaotic regions along with first and second order islands.
\item [\hbox to 0.32in {D:}] For $0.3\lesssim\alpha\lesssim0.5$ or $0.1\lesssim e\lesssim0.2$  chaos wipes out the second order islands. First order islands persist, but gradually diminish in size. Averaging and first order perturbation theory are still useful outside the chaotic regions.
\item[\hbox to 0.32in {E:}] For $\alpha\gtrsim0.5$ and $e\gtrsim0.2$ phase space is mostly chaotic, except for very small resonant islands.
\end{description}

Second order perturbation theory is important in regions B and C,
where $0.05\lesssim\alpha\lesssim0.3$ and $e\gtrsim0.1$. We are not
aware of any objects whose parameters are known that fall in regions B
and C, but it is possible that as more $\alpha$ values are ascertained
such objects will be identified. It seems likely that bodies
inhabiting a second order spin-orbit resonance do exist.

To illustrate the results of our perturbative calculations, let us
first consider a hypothetical body roughly at the boundary of regions
B and C; we choose $\alpha=e=0.2$ and consider its surface of section
in detail in Fig.~\ref{fig5}.

Fig.~\ref{fig5}a shows the second order perturbative model (that is,
contours of the second order fast action) covering first and
second order primary resonances from $1:2$ through $2:1$. First order
islands will have stable equilibria at $\theta=0,\pm{\pi}$ if
$H_{k}(e)>0$, and at $\pm{\pi}/{2}$ for $H_{k}(e)<0$. Here the $1:2$
resonance is of the latter type because $H_1(e)<0$, whereas the $1:1$,
$3:2$, and $2:1$ resonances are of the former type. For second order
resonances, which here are $3:4$, $5:4$, and $7:4$, the situation is
more complicated, but the same principle holds.
  
Comparison of Fig.~\ref{fig5}a with \ref{fig5}b shows excellent
agreement between second order perturbation theory model and numerical
integrations through most of phase space. But perturbation theory
naturally fails in the chaotic regions. Our perturbative model also
fails to reproduce the chain of secondary islands surrounding the
synchronous island in Fig.~\ref{fig5}b; our model traces contours
through the whole chain. We will discuss this issue in more detail
below.

In Figs.~\ref{fig5}c \& \ref{fig5}d we show respectively the curves
generated by the first order theory and by averaging. The averaging
technique reasonably approximates the $1:1$ and $3:2$ zones at these
parameters. First order perturbation theory improves on averaging in
that asymmetries in the islands, unaccounted for by the
averaging, now become apparent. But second order resonances are
not reproduced.

We now consider another hypothetical body, this time in region E.  We
choose $\alpha=e=0.65$ and show results for it in Fig.~\ref{fig6}.
The averaging contours in Fig.~\ref{fig6}a show resonance overlap, and
large-scale chaos is expected. However, as Fig.~\ref{fig6}b shows,
chaos does not completely pervade phase space and many small resonant
islands survive. Fig.~\ref{fig6}c shows that second order
perturbation theory can partially recover the $1:2$ islands even deep
inside region E.

\subsection{Particular objects}
 
\subsubsection{The Moon}

As our closest neighbor in space, the Moon has been a natural stimulus
and indeed testing ground for many theories of celestial
mechanics. Its occupation of the synchronous $1:1$ state has allowed
analysis of such tidal locking to be well studied for the many years
prior to the confirmation that the same resonance (with respect to the
corresponding parent planet) was shared by most other satellites in
the solar system.

Since the Moon has a relatively low $e$ and $\alpha$, putting it in
region~A, we anticipate that even first order perturbation theory
should provide a good match to the real system. Indeed, in
Fig.~\ref{fig7} the second order perturbative and numerical surfaces
of section are indistinguishable. The synchronous island and the $3:2$ islands are the only commensurabilities evident in the range plotted. We observe that in this instance there is scarcely any chaos bordering the separatrix.

\subsubsection{Mercury}

Let us investigate whether the agreement is comparable in the case of
Mercury which has a more elongated orbit than that of the Moon, and is
trapped in the $3:2$ resonance. It remains the only known example of
a non-synchronous primary resonance in our solar system. Mercury
spins on its axis once every $59.65$ days while taking roughly $1.5$
times as long to complete a single orbit in $87.97$ days. The current
eccentricity is $e=0.206$, but \citet{correiaandlaskarnature} show
that chaotic evolution of Mercury's orbit is capable of driving $e$ to
$\simeq0.45$. The asphericity is $\alpha=0.0187$ putting Mercury in
region~A.

In Fig.~\ref{fig8} we illustrate the dynamics in the neighborhood of
$3:2$ and $5:4$ resonances. The second order perturbative and
numerical surfaces of section are almost indistinguishable. The
second order $5:4$ resonance -- though it exists -- is tiny compared to
the first order $3:2$ resonance, and this is typical of region~A. We
found that for Mercury's low asphericity, the width of the $3:2$
resonance is remarkably insensitive to $e$. Also, the $5:4$ islands
remain tiny even at $e\simeq0.5$. Thus it is very improbable that
Mercury would ever have been trapped in the $5:4$ or other
second order resonance. The aforementioned proximity of Mercury's
rotation rate to the nominal location of the $5:4$ resonance seems to
be a coincidence.

\subsubsection{Hyperion} 

The chaotic nature of Hyperion's rotation has already been mentioned
in Sect.~\ref{Introduction}. Its large asphericity
$\alpha\simeq0.89$ together with eccentricity $e=0.1236$ puts Hyperion
off the scale of Fig.~4 but at a location that would correspond to a position deep in region E. As evident from e.g. Fig.~1 of \citet{blacketal},
the phase space is largely chaotic but has some small resonant
islands. Our second order perturbative model recovers the $5:2$
resonant islands, as shown in Fig.~\ref{fig9} but does not succeed for
islands below $\dot{\theta}/n\approx2.5$. (In this example, we
increased $k_{\rm max}$ in Eq.~(\ref{primaryH}) from 4 to 6, because of
the comparatively large perturbation.)

\subsubsection{Enceladus}

The saturnian satellite Enceladus has a moderately high asphericity
$\alpha=0.336$ but very low eccentricity $e=0.0045$, thus putting it
in region~A. It exhibits a secondary resonance, a phenomenon not
allowed for in our perturbative model. Fig.~\ref{fig10} shows our
results for the $1:1$ resonance in Enceladus. With relation to the
secondary islands and their separatrix, we find that our perturbative
model fails to resemble these lobes; instead concentric rings
intersect these areas. On the other hand, \citet{wisdom2004} shows
that the secondary resonances can be reproduced by an averaging model
specifically designed for secondary resonances. To facilitate
comparison, Fig.~\ref{fig10} has been chosen to have the same scales
as Wisdom's Fig.~2a.

We will now digress briefly to explain the difference between Wisdom's
approach and ours.

\subsection{Secondary Resonance Dynamics}

Let us return for a moment to the main Hamiltonian Eq.~(\ref{primaryH}).
Considering only the synchronous resonance, the Hamiltonian becomes
\begin{equation}
\label{synch}
H(p_1,\,p_2;\,q_1,\,q_2) =  \frac{{{p_1}}^2}{2} + {p_2} - 
\frac{\alpha^2}{4}\, \left(1-\frac{5\,e^2}{2}+\cdots\right)\,
\cos{(2\,{q_1} - 2\,{q_2})}.
\end{equation}
Recall that we have taken
\begin{equation}
\label{flynn_sol}
H_0(p_1,\,p_2;\,q_1,\,q_2) =  \frac{{{p_1}}^2}{2} + {p_2}, 
\end{equation}
as the unperturbed Hamiltonian and $\alpha^2/4$ as the perturbing
parameter. The action-angle variables of $H_0$ are simply are
$p_1,\,p_2,\,q_1,\,q_2$. On the other hand Wisdom adopts
\begin{equation}
\label{flynn_solb}
H_0^{W}(p_1,\,p_2;\,q_1,\,q_2) =  \frac{{{p_1}}^2}{2} + {p_2}- 
  \frac{\alpha^2}{4}\,\cos{(2\,{q_1} - 2\,{q_2})}.
\end{equation}
as the unperturbed Hamiltonian and $e$ as the perturbation. If this is done, the synchronous resonance appears already in the unperturbed dynamics, and then through perturbation theory the secondary resonances can be recovered. The complication is that the action-angle variables of $H_0^{W}$ are no longer
$p_1,\,p_2,\,q_1,\,q_2$ but non-elementary functions of them; moreover the
perturbation must be expressed in terms of these new action-angles. Fortunately in the case of Enceladus, because of the small
eccentricity, averaging is adequate and for averaging it is only
necessary to extract one key term in the perturbation.

Developing first or second order Lie transform perturbation theory for
secondary resonances is in principle possible but is a much more
difficult task because of the complicated nature of the action-angle
variables required.

Secondary resonances are themselves part of a hierarchy which may
extend down to smaller and smaller scales. As an illustration, we
show a numerical surface of section for the saturnian satellite
Pandora in Fig.~\ref{fig11}. Zooming in successively, we see primary,
secondary, tertiary, and quaternary resonant islands.

\section{Conclusions}\label{Conclusions}

We have applied the technique of Lie transform perturbation theory to
the planar spin-orbit problem, to second order. The full perturbative
expressions are too long to include in the paper, but fortunately the main features of spin-orbit dynamics have analogs in the simpler
problem of a driven double pendulum, which allows us to explain our
perturbative method without excessive algebra.

We have compared our perturbative model to numerical integrations for
various values of the asphericity parameter $\alpha$ and the orbital
eccentricity $e$. If at least one of these is small 
($\alpha\lesssim0.05$ or $e\lesssim0.1$) then only first order
resonances are important and first order perturbation theory is adequate;
Mercury and the Moon lie in this regime. If the tidal perturbations
become too large ($\alpha\gtrsim0.5$ and $e\gtrsim0.2$) then the spin
becomes chaotic and perturbation theory fails altogether; Hyperion is the
best-known example. However, for intermediate perturbations
($0.05\leq\alpha\leq0.3$ and $e>0.1$) second order resonances are
possible and second order perturbation theory can reproduce them.

Our perturbative model is limited to primary resonances. In
particular, Enceladus is thought to be in a secondary $3:1$ resonance
around the primary synchronous resonance. Our model smooths over
secondary resonances, putting Enceladus in an ordinary primary
resonance. However, \citet{wisdom2004} has shown how to recast the
problem so that the secondary resonance can be correctly reproduced in
perturbation theory. Extending Wisdom's model to second order is
possible in principle, but very complicated.

So far, none of the objects whose asphericities may be found in the
literature fall in the region of $\alpha,\,e$ where second order
perturbation theory is particularly interesting, that is, regions B
and C of Fig.~\ref{fig4}. But there is no doubt that objects in the
interesting parameter range do exist, and the possibility of objects
being locked in second order resonances remains open.

\acknowledgments This research was financially supported by the UK
Particle Physics and Astronomy Research Council. We thank Nick Cooper, Carl
Murray, John Papaloizou, Jack Wisdom, and the referee for comments that improved this
manuscript.

\appendix

\section{The driven pendulum}

The Hamiltonian
\begin{equation}
H(p,\,q,\,t) = \frac{p^2}{2} - \alpha\,\cos{q} - \beta\,\cos{(q-t)}
\end{equation}
has been widely studied in the literature on the transition to chaos,
e.g. \citet{escande}. Several possible physical realizations of this
Hamiltonian are known, typically involving electric or magnetic
fields. Here we present another physical interpretation, which is very
simple to visualize and is intuitively analogous to the the spin-orbit
problem.

\begin{figure}[h]
\epsscale{0.5}\plotone{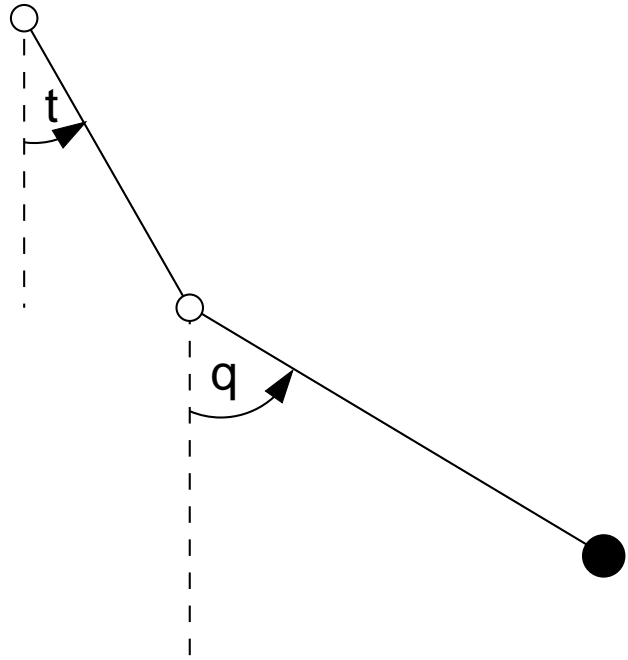}
\caption{Mechanics of the driven double pendulum. The open circles correspond to weightless hinges while the filled circle indicates the mass.
\label{fig2}}
\end{figure}

\begin{figure}[ht]
\epsscale{0.5}
\plotone{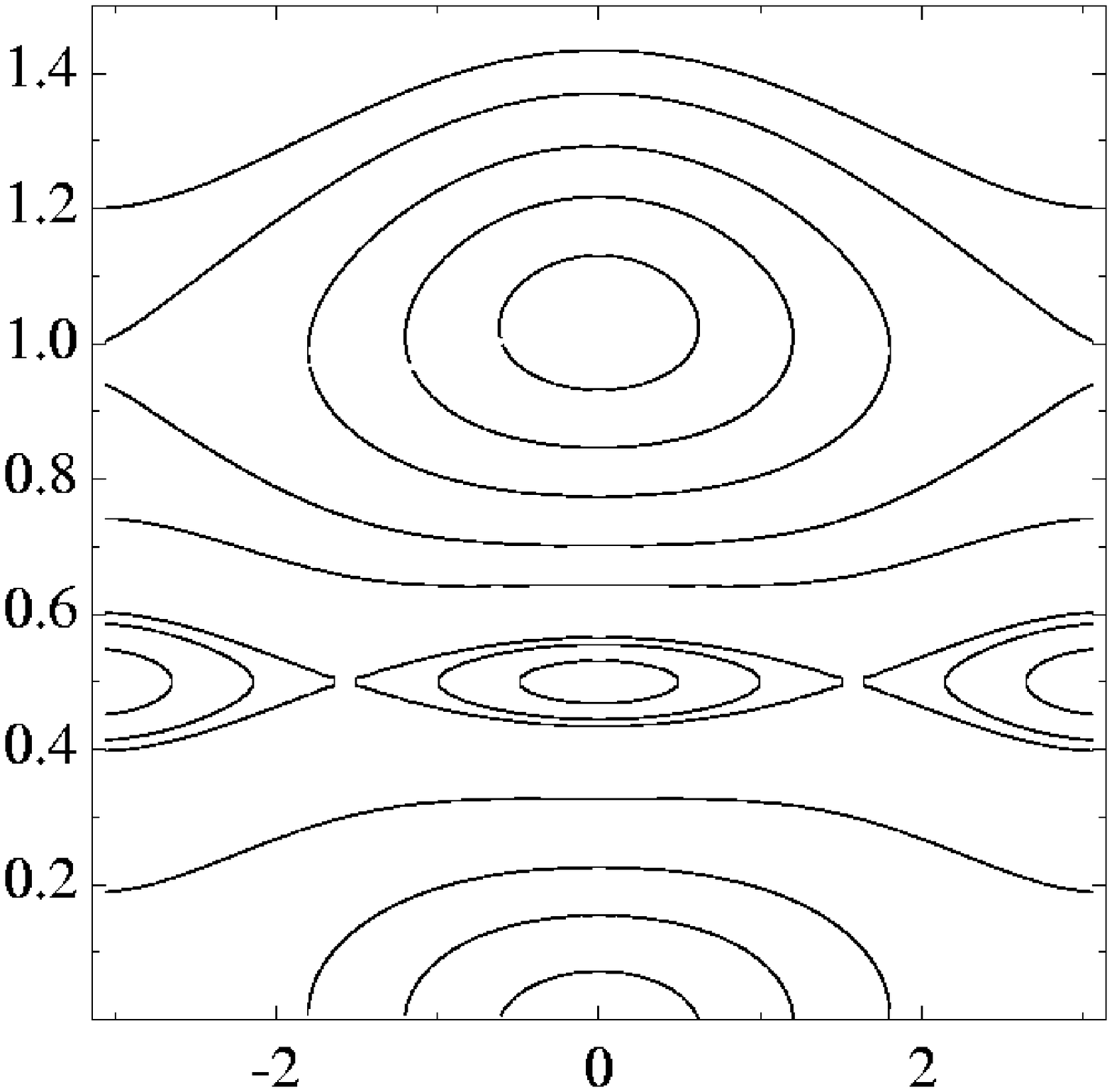} \par
\plotone{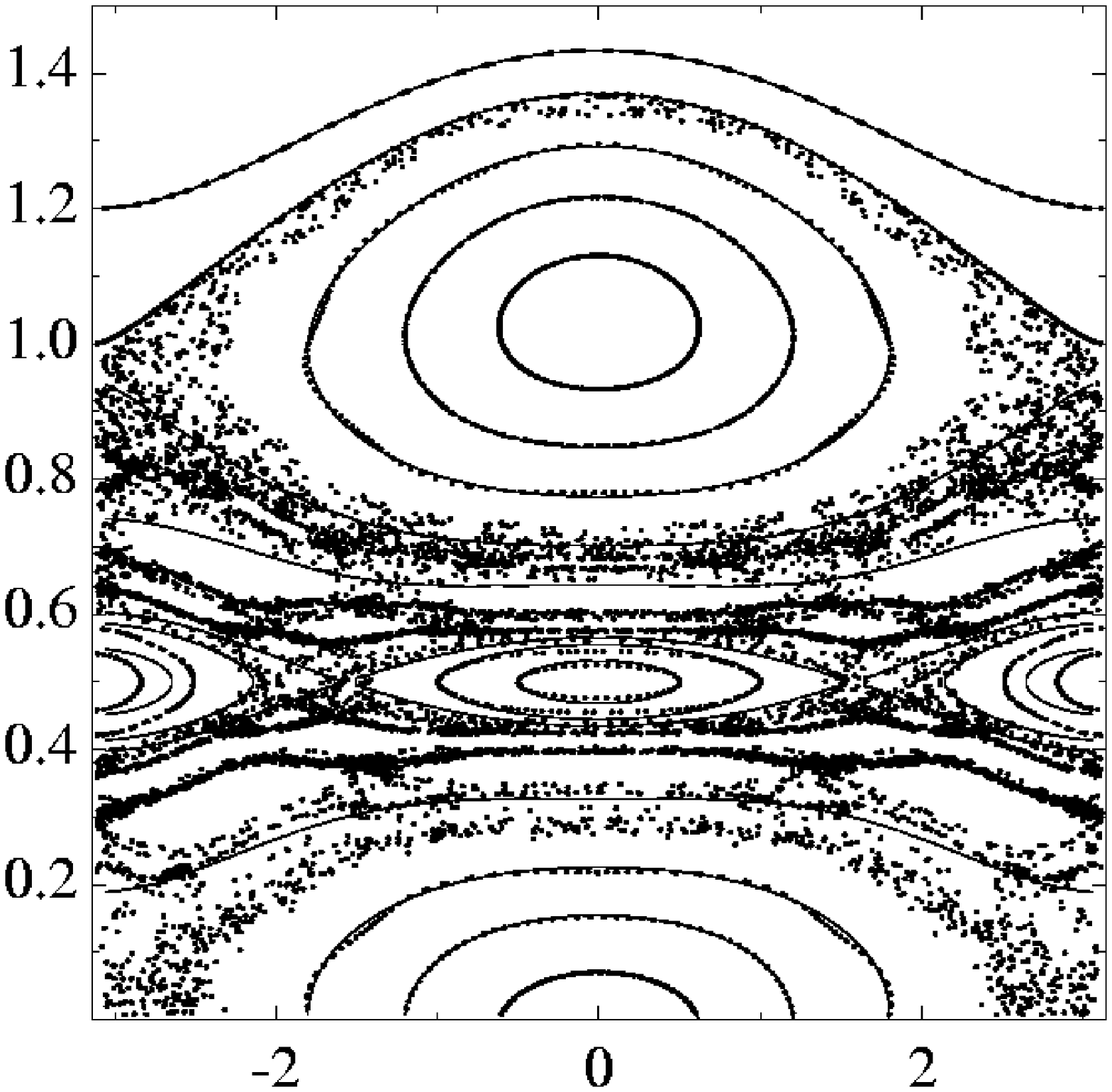}
\caption{Upper panel: Contours of the fast action (slice chosen at
$q_2=0$) for the driven pendulum with $\alpha=\beta=0.03$. Lower
panel: Numerical surface of section for the driven pendulum for the
same parameters. The second order $1:2$ resonance is clearly visible
centered on $p_1=0.5$. Also evident are the first order $0:1$ and
$1:1$ states. The horizontal axis measures $q_1$ while the vertical
axis measures $p_1$.
\label{fig3}}
\end{figure}

\begin{figure}[h]
\epsscale{0.8}\plotone{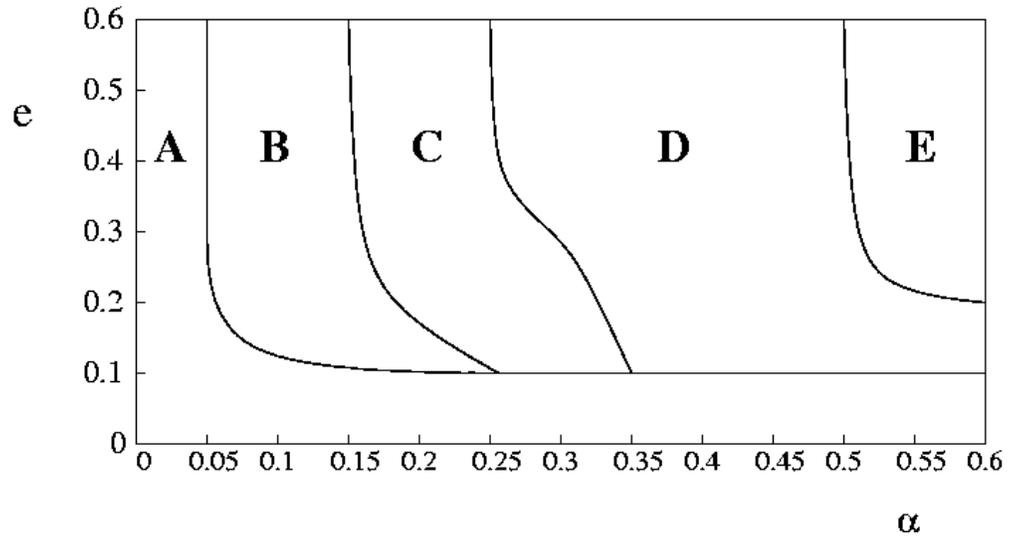}
\caption{Parameter regimes of the spin-orbit problem. A:~non-resonant
  spins and first order resonant islands. B:~similar to A, but with
  second order islands also present. C:~like B but with significant
  chaos. D:~second order islands overrun by chaos, first order
  islands remain. E:~large-scale chaos with tiny or no resonant
  islands.
\label{fig4}}
\end{figure}

\begin{figure}
\epsscale{0.35}
\plotone{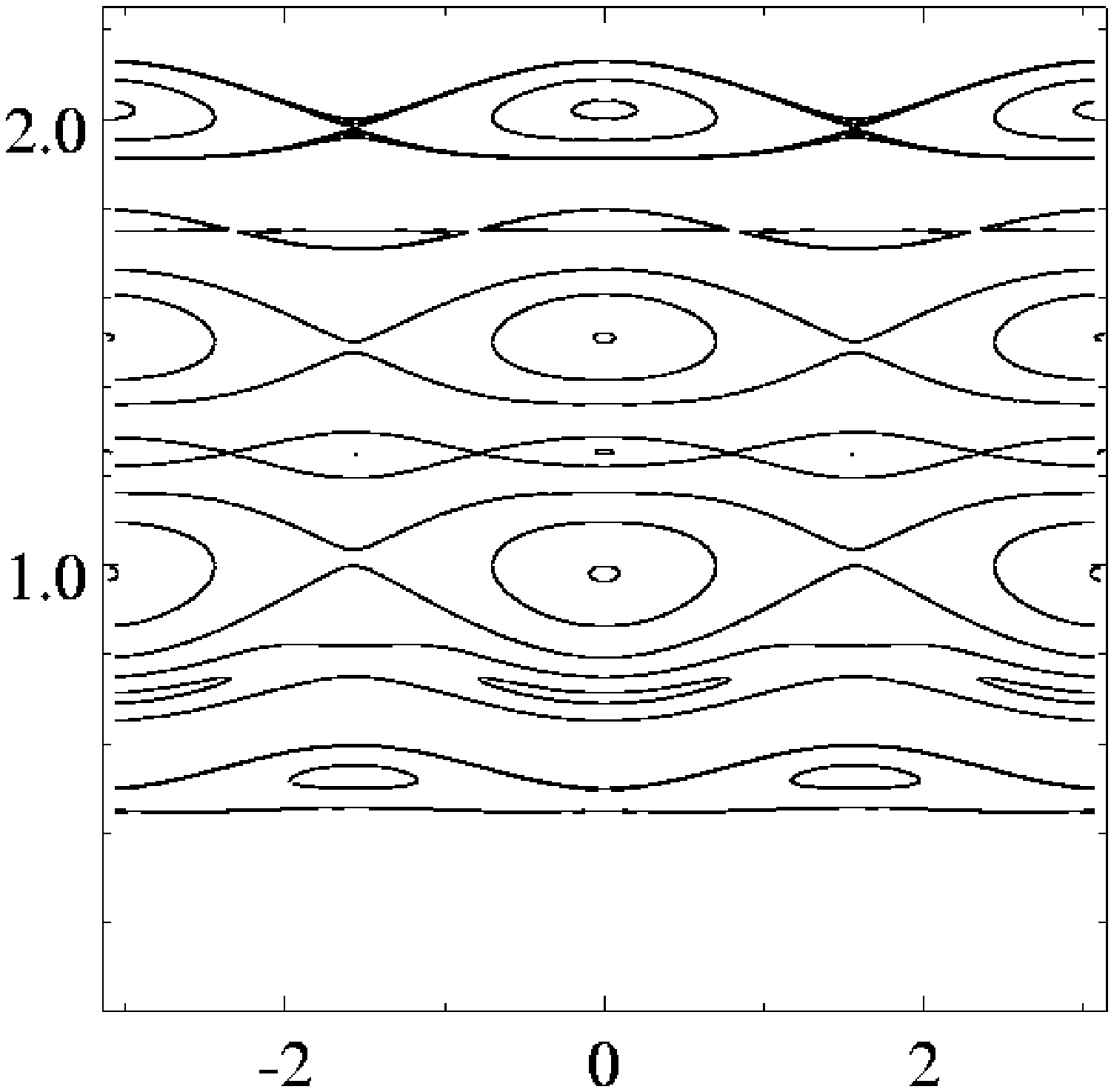}\plotone{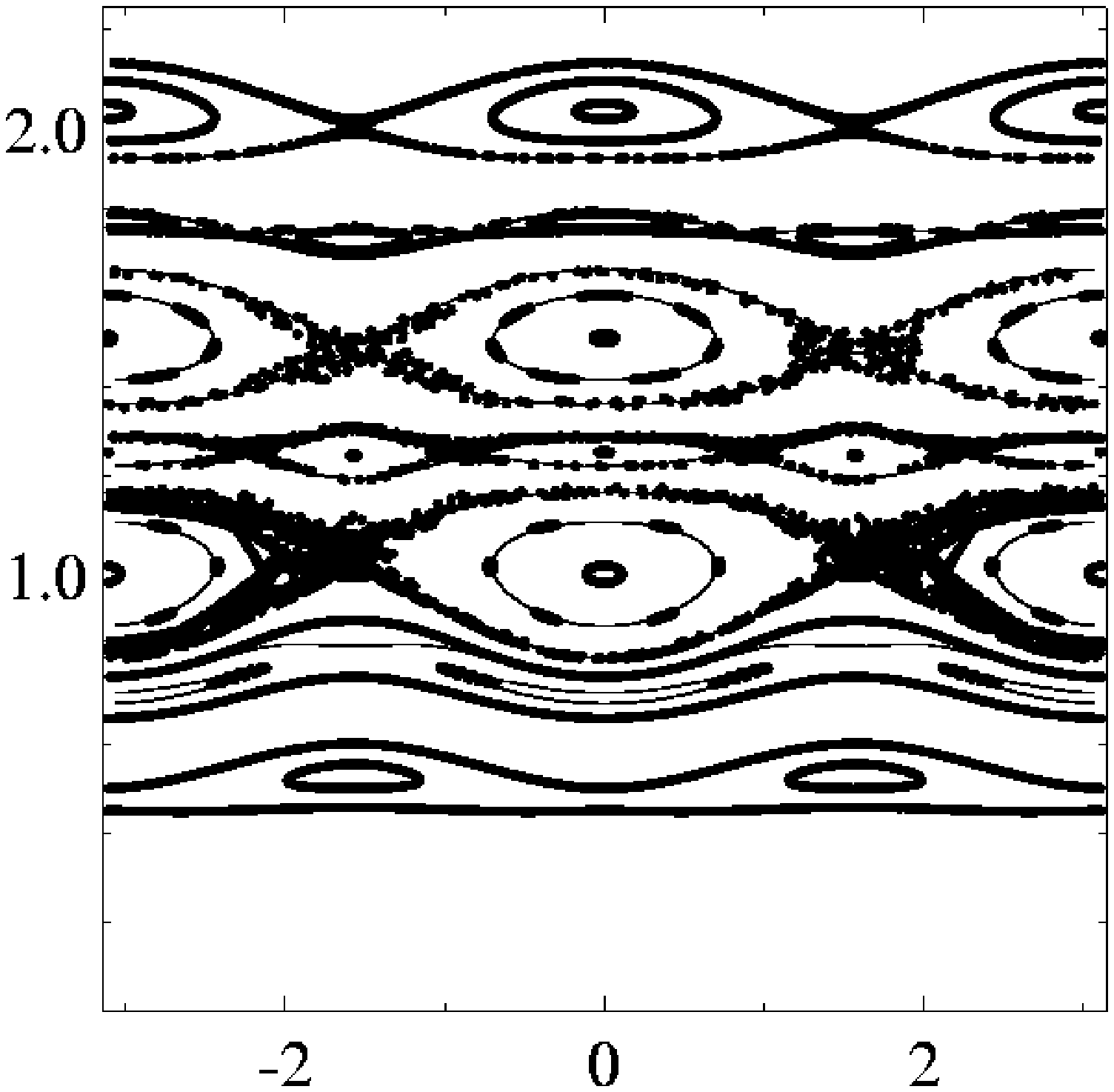} \par
\plotone{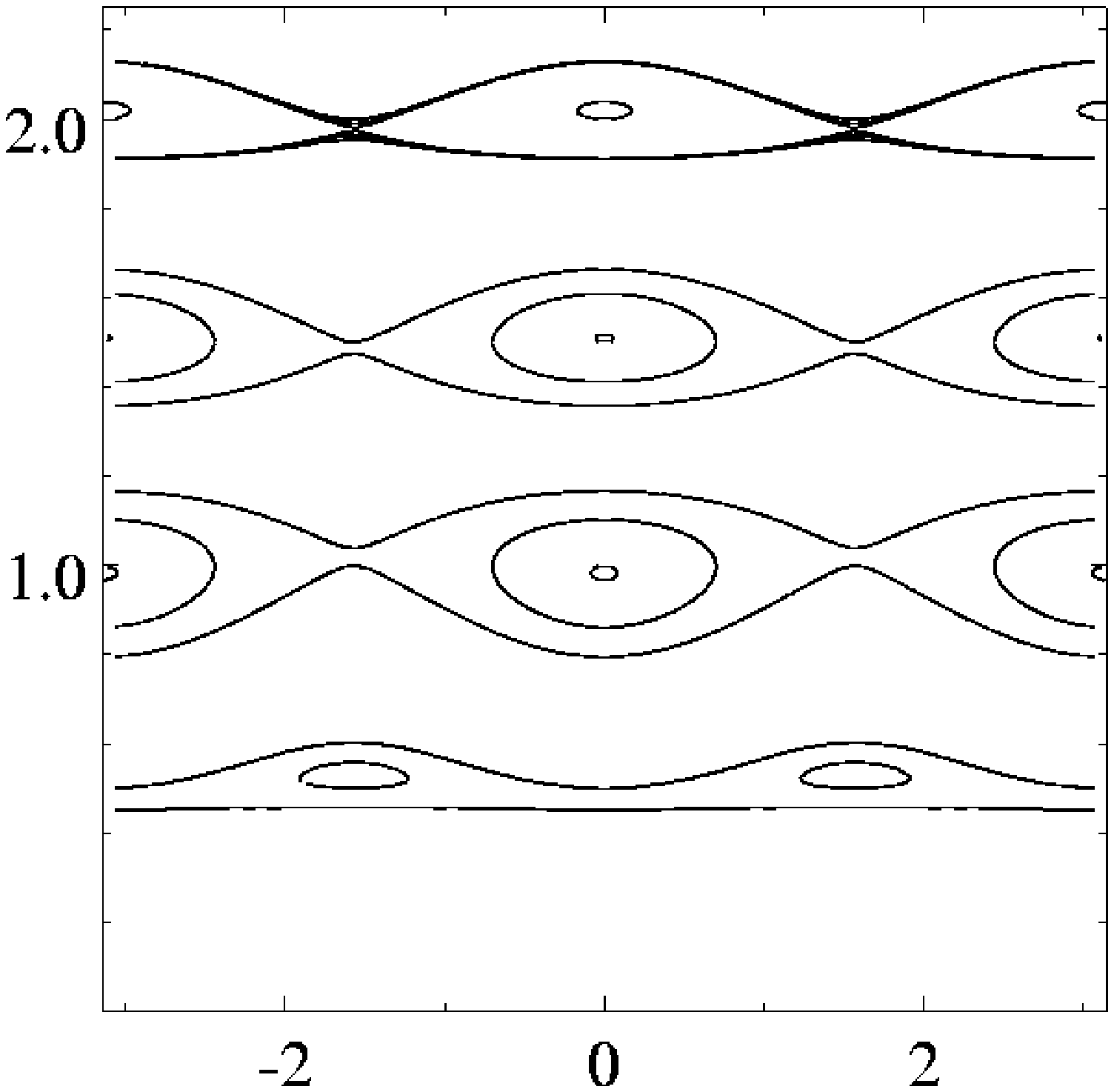}\plotone{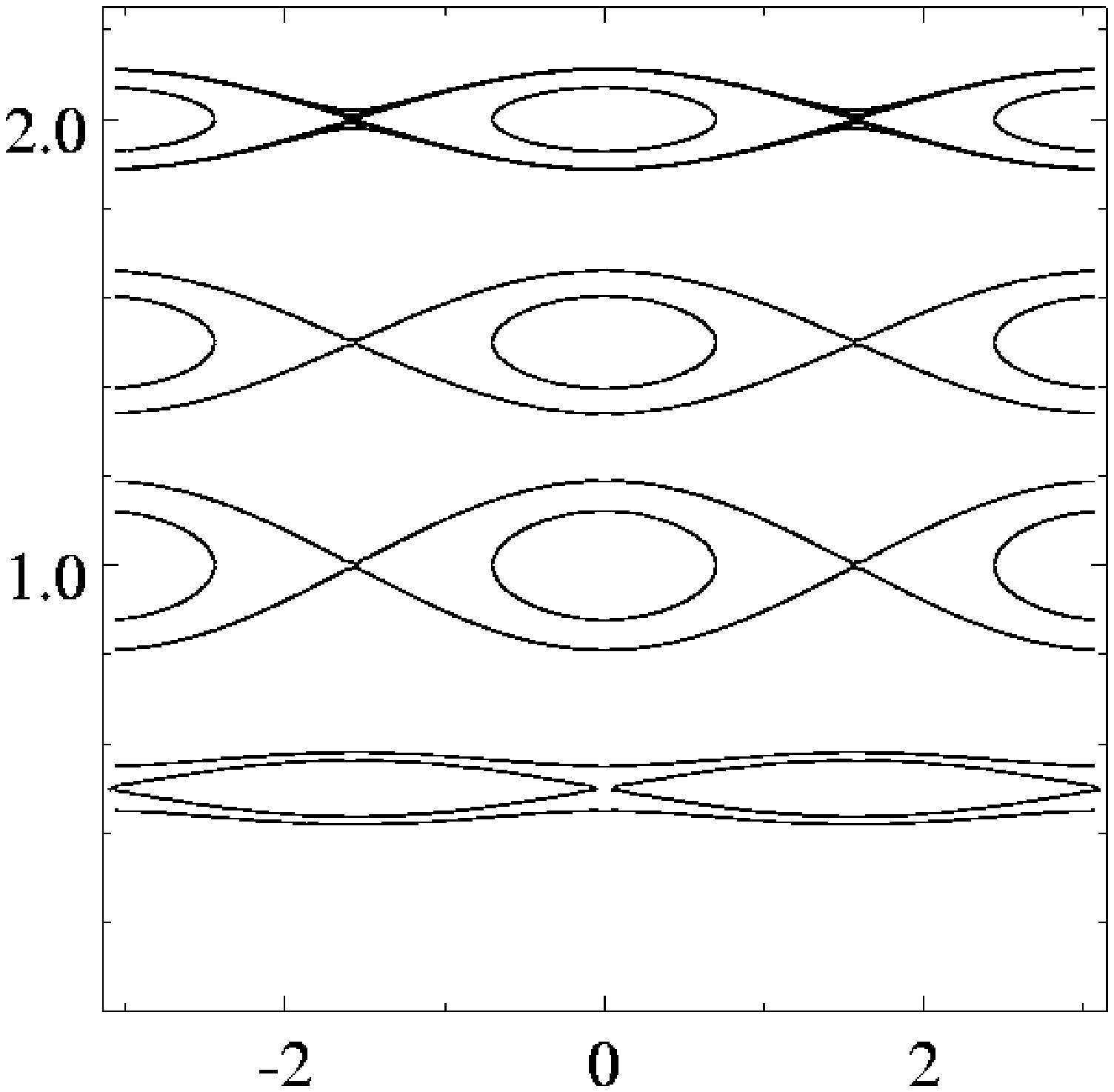}
\caption{(a)~Upper left panel: Surface of section, analogous to
Fig.~\ref{fig3} but for the spin-orbit problem, illustrating the
second order perturbation theory for the $1:2,\, 3:4,\, 1:1,\, 5:4,\,
3:2,\, 7:4$ and $2:1$ resonances for $\alpha=0.2;\, e=0.2$. (b)~Upper
right panel: A numerical surface of section for the same parameters.
(c)~Lower left panel: First order result. (d)~Lower right panel:
Averaging result. In all our surfaces of section the horizontal axis
measures $q_1=\theta$ and the vertical axis measures $p_1=\dot{\theta}/n$.
\label{fig5}}
\end{figure}

\begin{figure}
\epsscale{0.35}
\plotone{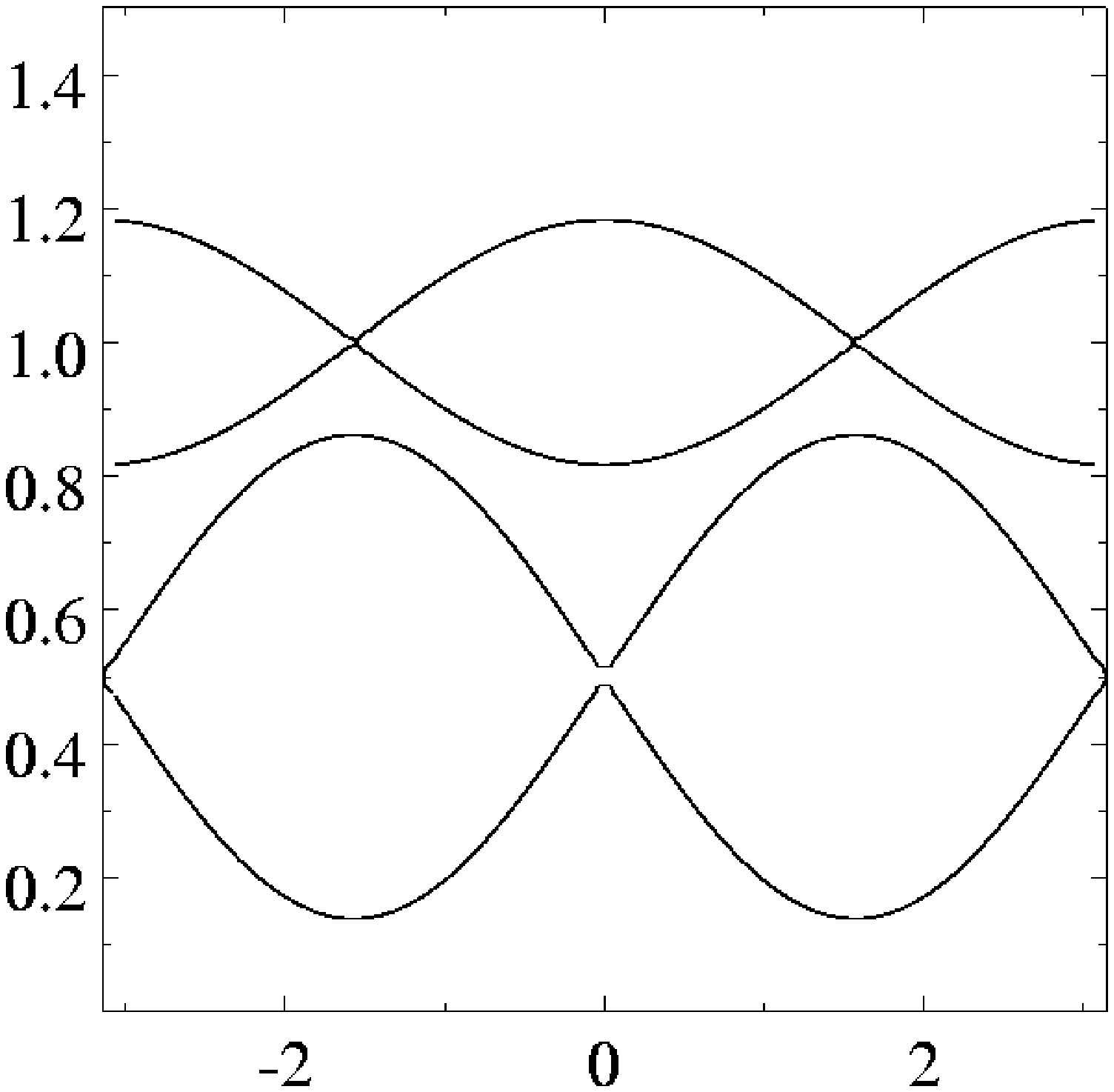} \par
\plotone{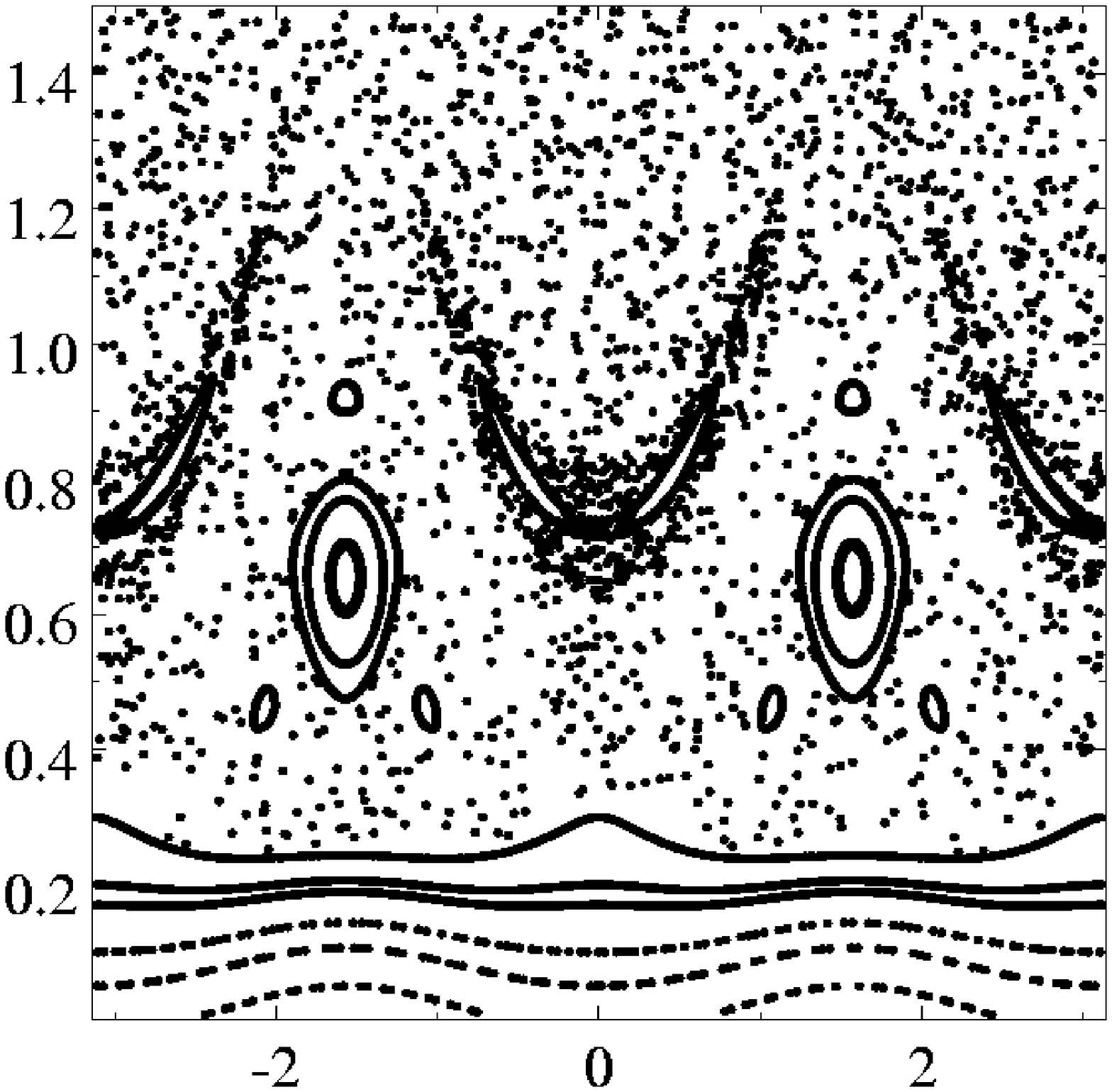} \par
\plotone{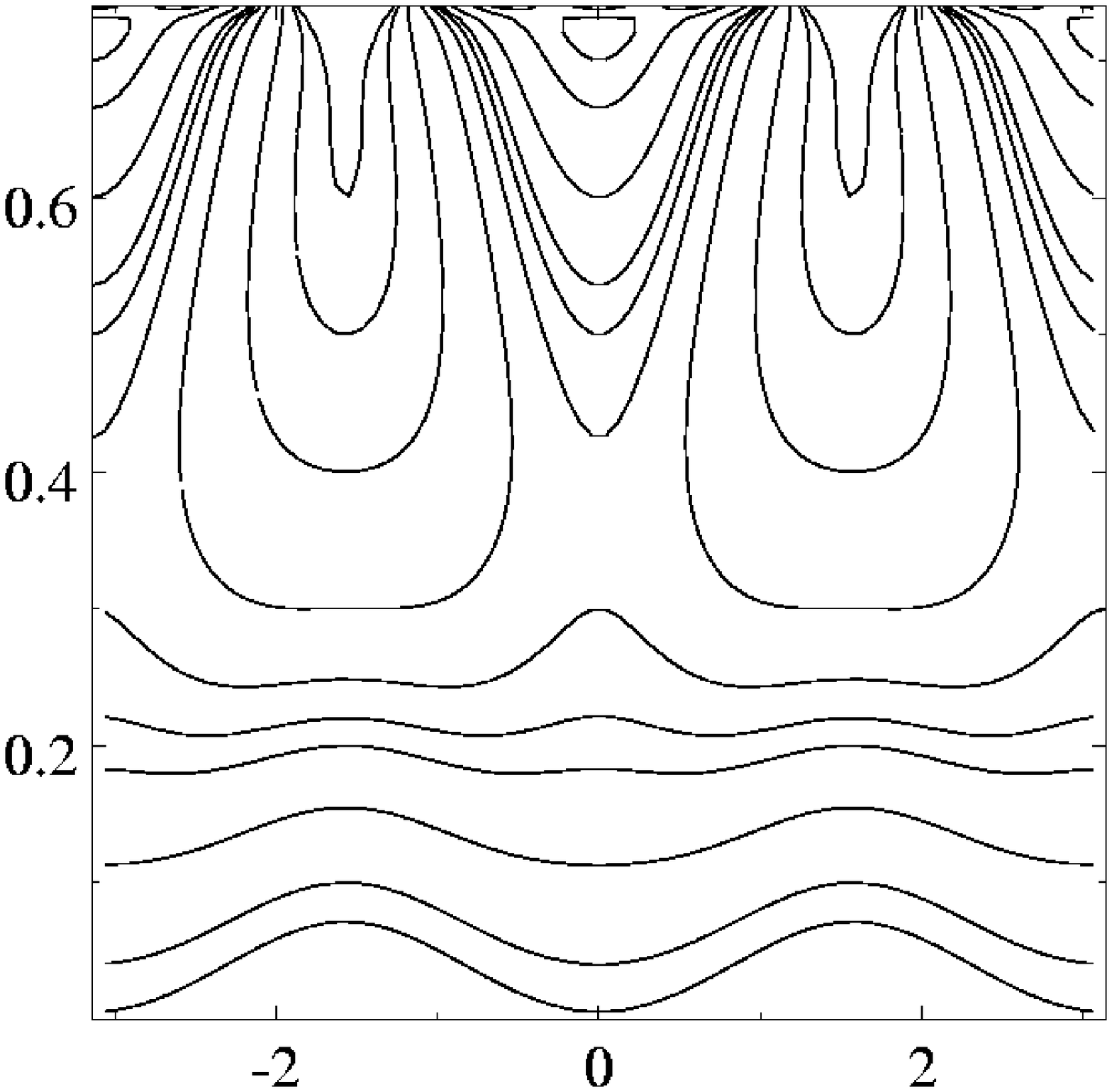}
\caption{Surfaces of section for $\alpha=e=0.65$. (a)~Upper panel:
Contours from averaging, illustrating overlap between the $1:2$ and
$1:1$ resonances. (b)~Middle panel: Numerical surface of section showing
large-scale chaos, but with numerous resonant islands. (c)~Lower panel:
Second order perturbative result partially recovering the $1:2$
islands. Note the change in scale from (a) and (b). The horizontal axes measure $q_1=\theta$ and the vertical axes measure $p_1=\dot{\theta}/n$.
\label{fig6}}
\end{figure}

\begin{figure}
\epsscale{0.35}
\plotone{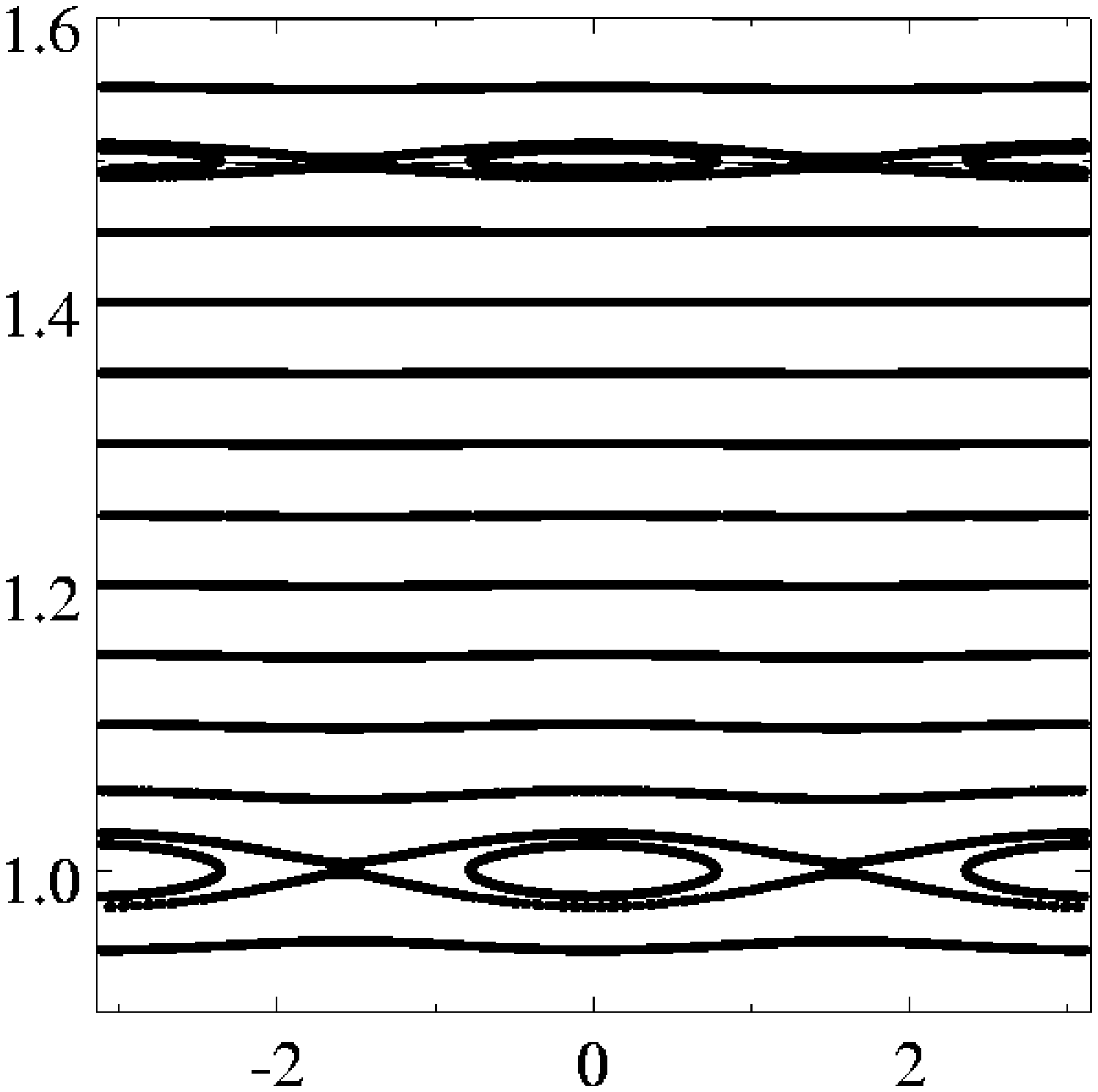} \par
\plotone{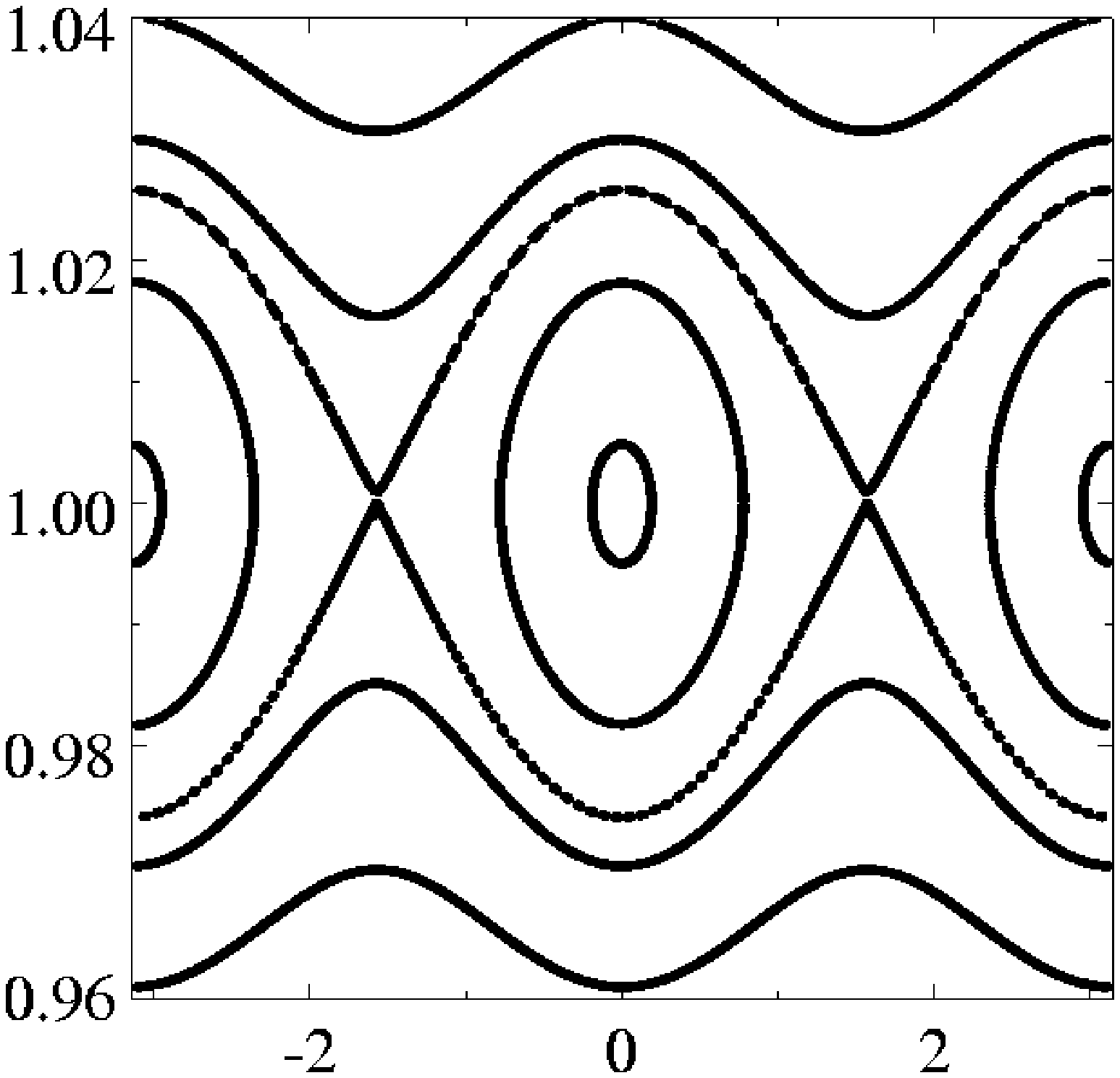}
\caption{(a)~Upper panel: Numerical surface of section (thick curves) for
the Moon superimposed on second-order perturbative contours. The
parameters are $\alpha=0.026;\,e=0.0549$. (b)~Lower panel:~Zoom of the
synchronous zone. The horizontal axes measure $q_1=\theta$ and the vertical axes measure $p_1=\dot{\theta}/n$.
\label{fig7}}
\end{figure}

\begin{figure}
\epsscale{0.35}
\plotone{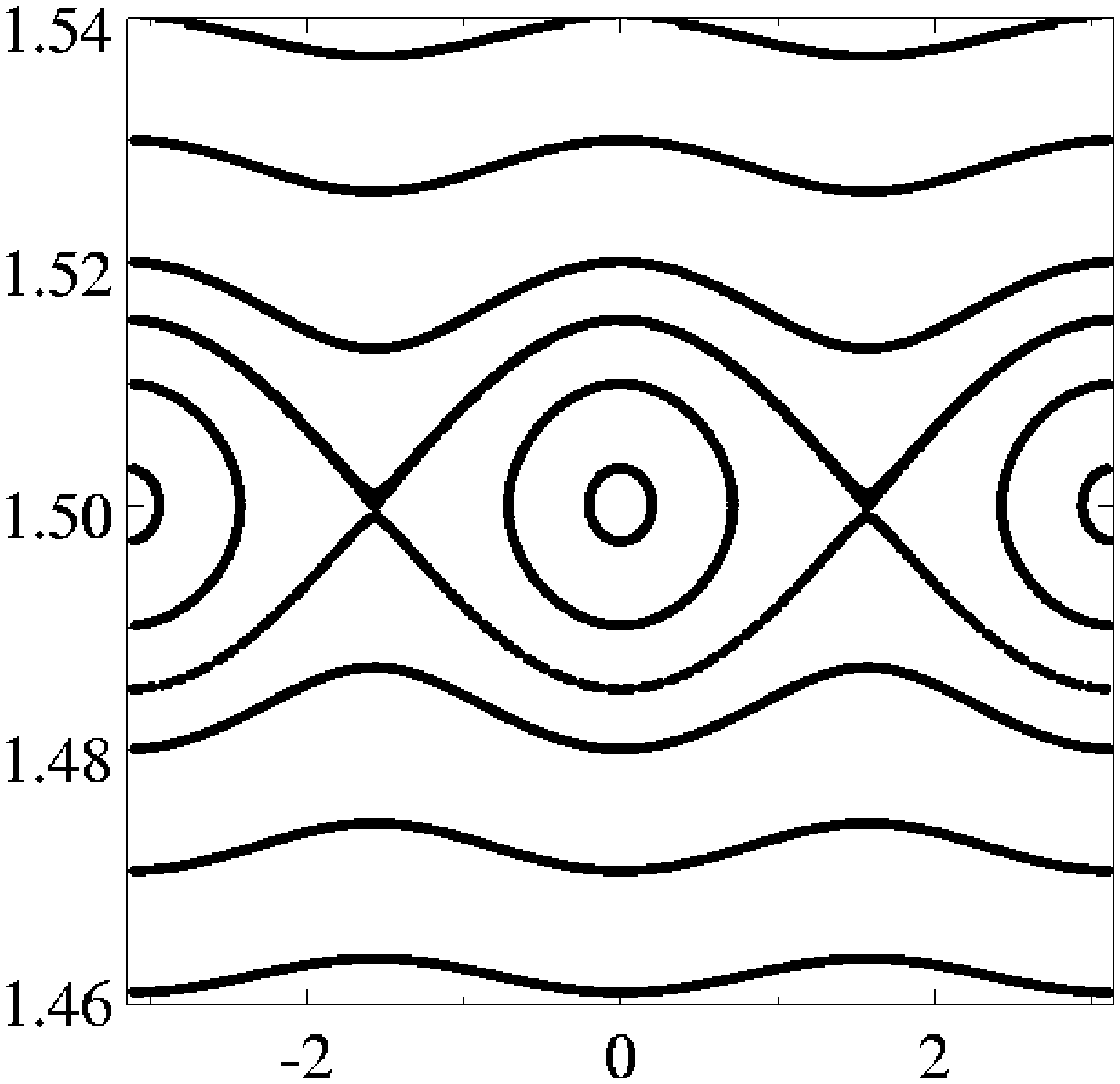} \par
\plotone{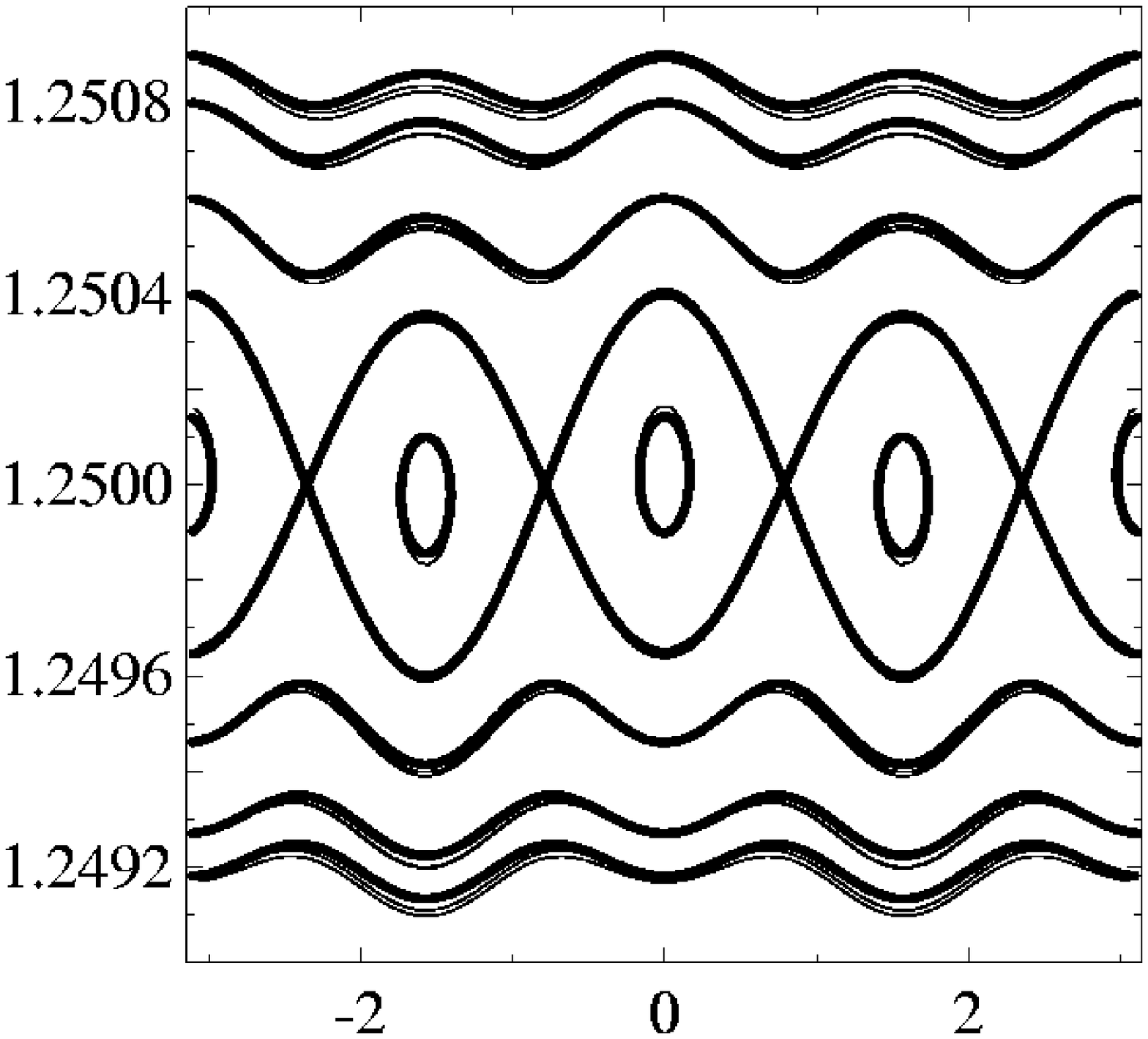}
\caption{(a)~Upper panel: Numerical surface of section for Mercury
(thick curves) superimposed on the second-order perturbative
contours. Only the region around the $3:2$ state is plotted. (b)~Lower
panel: Similar, but near the second-order $5:4$ resonance. Note the
change in scale from Fig.~\ref{fig8}a. The parameters are
$\alpha=0.0187;\,e=0.206$. The horizontal axes measure $q_1=\theta$ and the vertical axes measure $p_1=\dot{\theta}/n$.
\label{fig8}}
\end{figure}

\begin{figure}
\epsscale{0.35}
\plotone{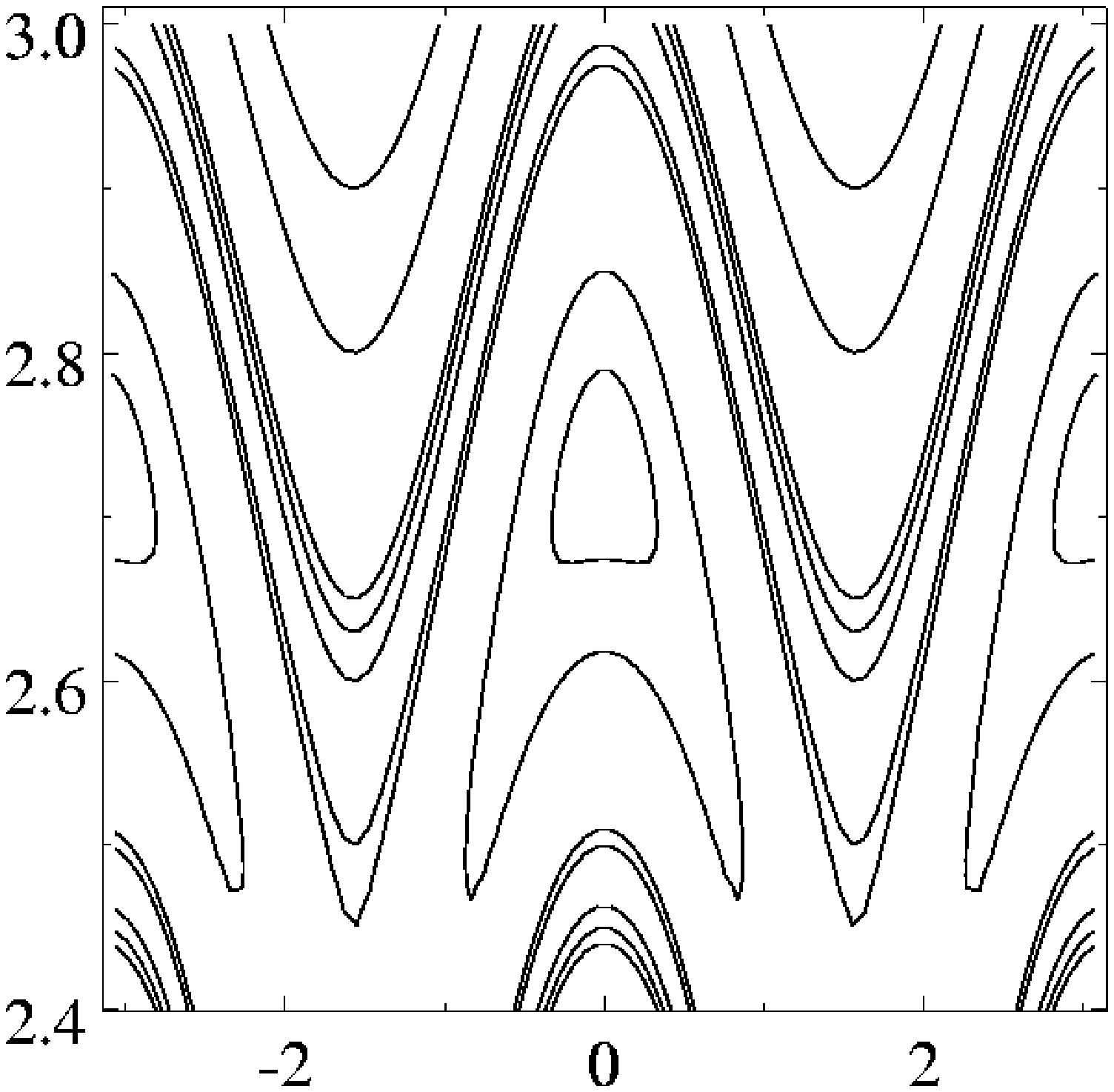} \par
\plotone{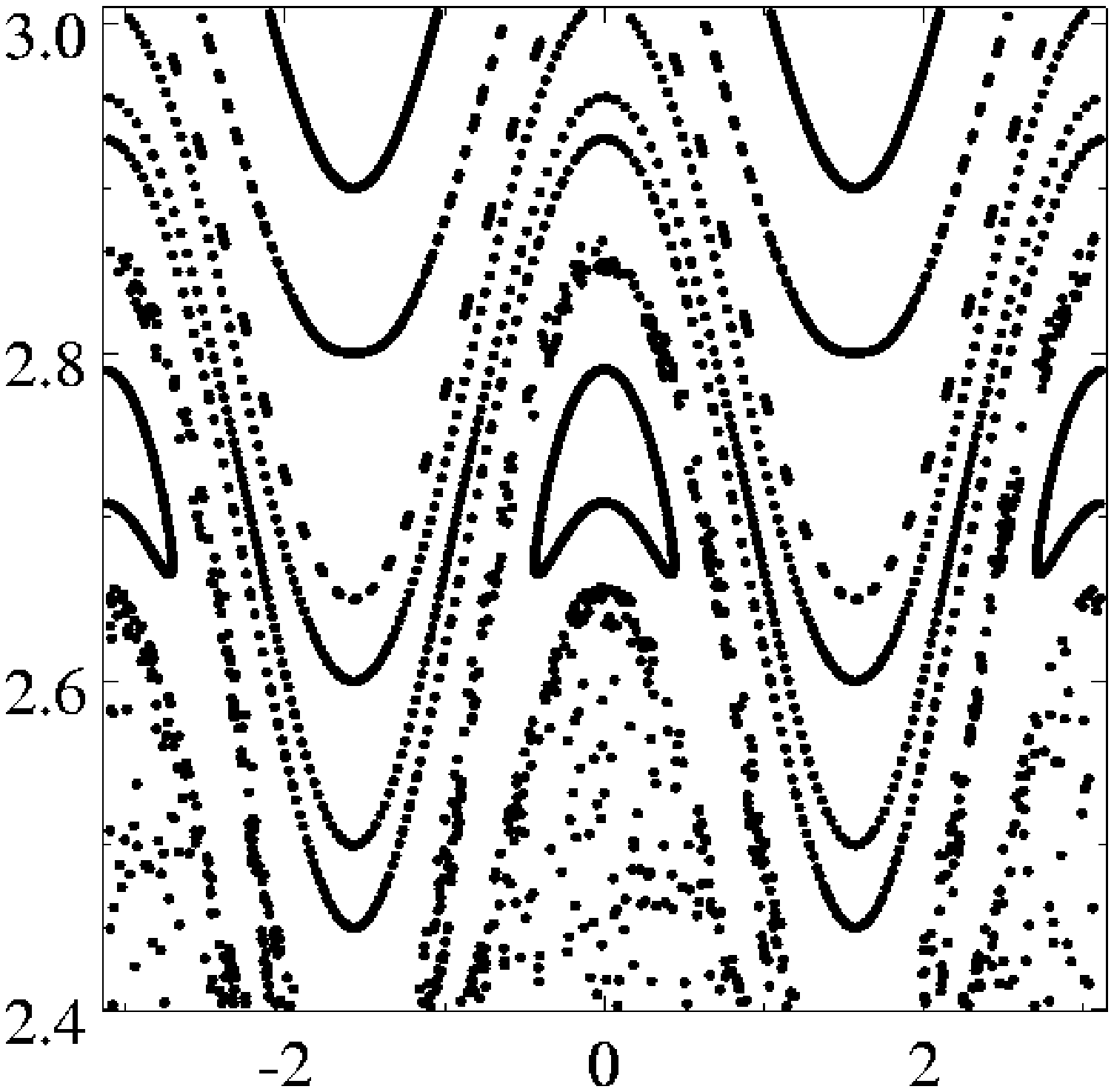}
\caption{The $5:2$ islands for Hyperion. The upper panel shows the
second-order perturbative result, while the lower panel is a numerical
surface of section. The parameters are $\alpha=0.89;\,e=0.1236$.
The horizontal axes measure $q_1=\theta$ and the vertical axes measure
$p_1=\dot{\theta}/n$.
\label{fig9}}
\end{figure}

\begin{figure}
\epsscale{0.35}
\plotone{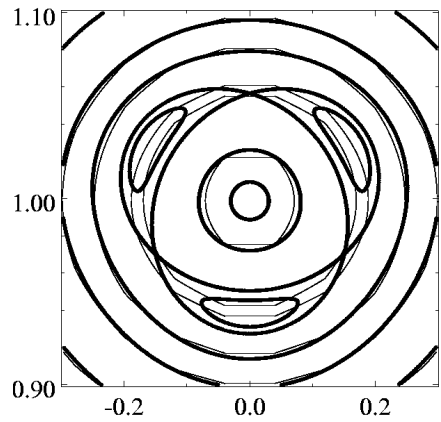}
\caption{Numerical surface of section (thick curves) for Enceladus
superimposed on the second-order perturbative contours. The parameters
are $\alpha=0.336;\,e=0.0045$. The horizontal axis measures $q_1=\theta$ and the vertical axis measures $p_1=\dot{\theta}/n$. The $3:1$ secondary islands encircling the synchronous libration island is not reproduced by our theory: in their place exist concentric circles. This figure is analogous to Fig.~2 from \citet{wisdom2004}. 
\label{fig10}}
\end{figure}

\begin{figure}
\epsscale{0.35}
\plotone{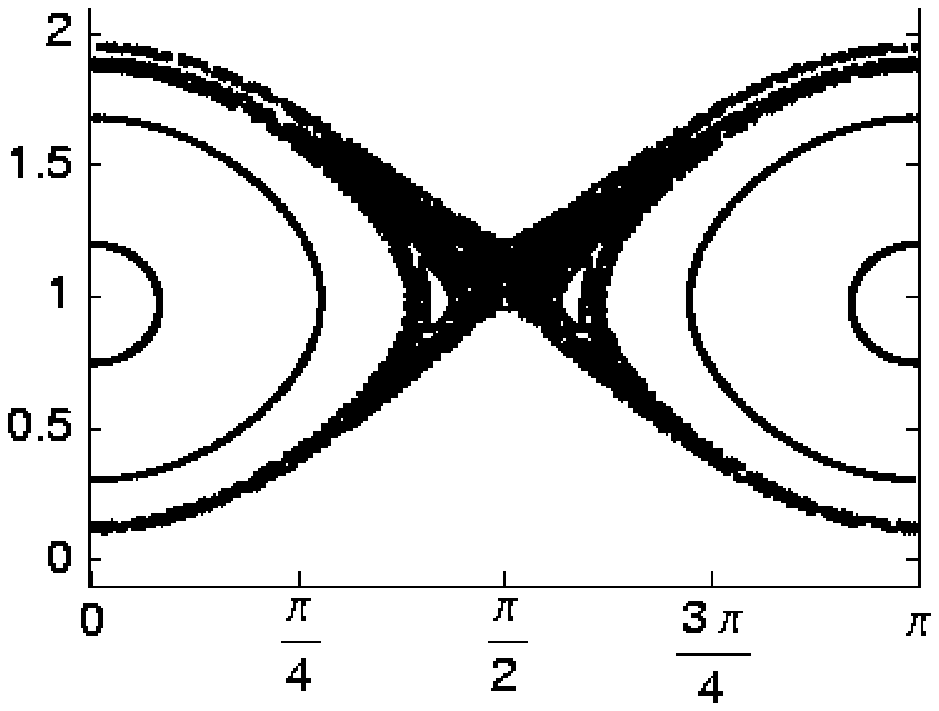} \par
\epsscale{0.35}
\plotone{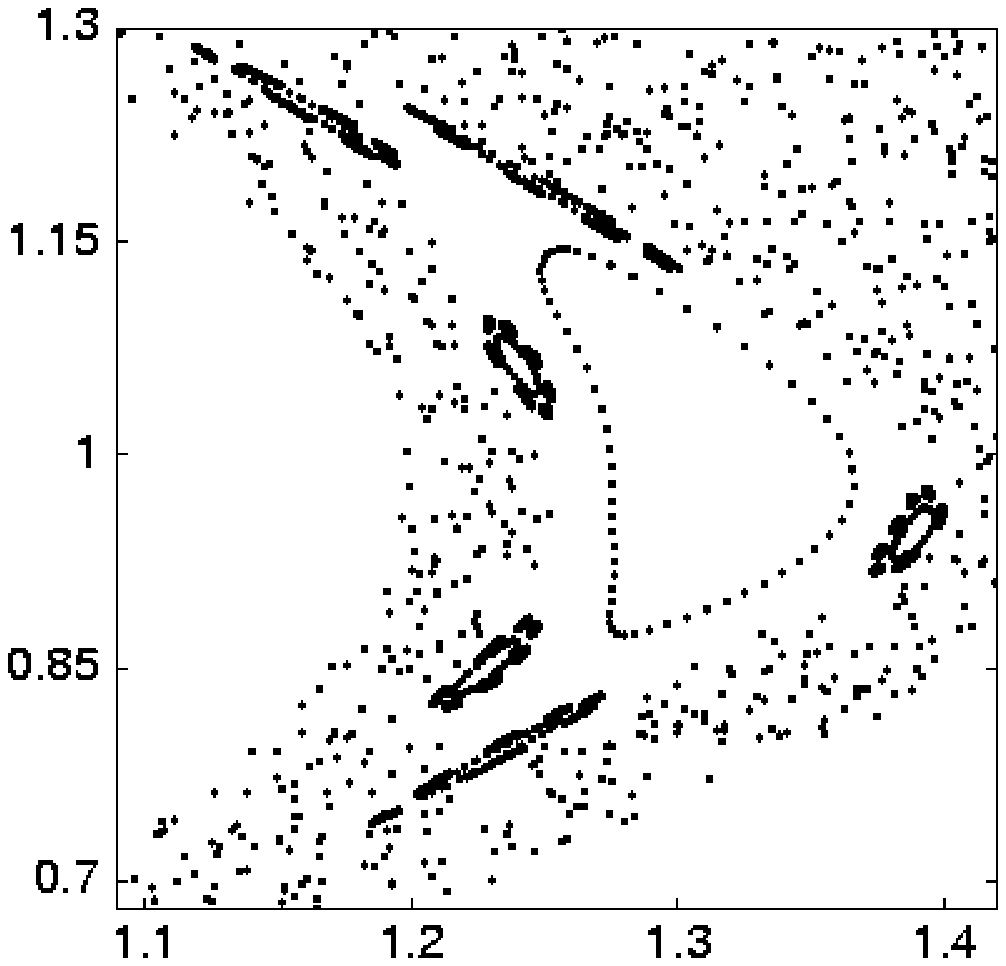} \par
\plotone{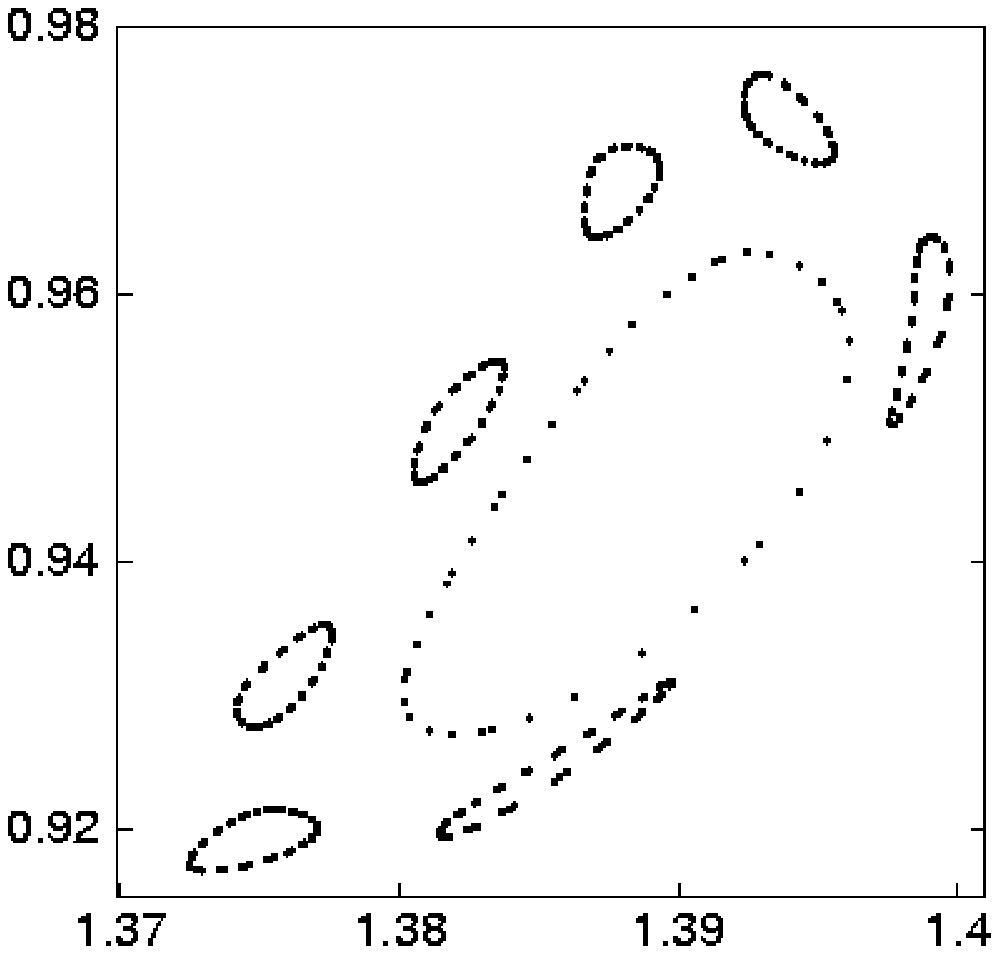}
\caption{Upper panel: Numerical surface of section for Pandora. The
parameters are $\alpha=0.89;\,e=0.0004$.  Middle panel: Zooming in on the
secondary island and its environs. Lower panel: Zooming in further in on
a tertiary isle, itself encircled by seven quaternary islets.
The horizontal axes measure $q_1=\theta$ and the vertical axes measure
$p_1=\dot{\theta}/n$.
\label{fig11}}
\end{figure}

Our physical system is illustrated in Fig.~\ref{fig2}. It is a double
pendulum with two light rods and hinges and a single bob at the end.
The inner pendulum (having length $b$, say) is made to circulate by an
external motor at unit angular frequency. The outer pendulum (having
length $l$, say) is free to librate or circulate. We can write the
coordinates of the bob as
\begin{eqnarray}
x &=&  l\,\sin{q} + b\,\sin{t} \cr
y &=& -l\,\cos{q} - b\,\cos{t}. 
\end{eqnarray}
The Lagrangian for the bob can be written as
\begin{equation}
L(q,\,\dot{q},\,t) = \frac{b^2}{2} + \frac{l^2\,\dot{q}^2}{2} + g\,l\,\cos{q}
+ l\,b\,\dot{q}\,\cos{(q-t)} + g\,b\,\cos{t} .
\label{lagrangian1}
\end{equation}
We now discard the first and last term. (The last term depends on
neither of $q,\,\dot{q}$ and hence will not contribute to the
equations of motion.) For algebraic convenience we also divide the
Lagrangian by $l^2$ and introduce the parameters
$\alpha=g/l,\,\beta=b/l$. As a result of all these changes
Eq.~(\ref{lagrangian1}) can be replaced by
\begin{equation}
L(q,\,\dot{q},\,t) = \frac{\dot{q}^2}{2} + \alpha\,\cos{q} + \beta\,\dot{q}\,\cos{(q-t)}
\label{lagrangian2}
\end{equation}
The interpretation of $\alpha,\,\beta$ is as follows: $\sqrt{\alpha}$ is
the natural frequency of the outer pendulum relative to the driving
frequency; $\sqrt{\beta}$ is the natural frequency of the outer pendulum
relative to the inner. Hence $\alpha=\beta$ corresponds to the inner
pendulum being driven at its natural frequency.

The Hamiltonian corresponding to $L$ in Eq.~(\ref{lagrangian2}) is
\begin{equation}
H(p,\,q,\,t) = \frac{\left(p-\beta\,\cos{(q-t)}\right)^2}{2} - \alpha\,\cos{q}.
\label{Hpend1}
\end{equation}
We can simplify (\ref{Hpend1}) using a canonical transformation.
Inserting the generating function\footnote{To convert to the notation of
\citet{goldstein} Sect.~9-1, read $-F_3$ for our $S$.}
\begin{equation}
S(p,\,Q) = p\,Q - \beta\,\sin{(Q-t)}
\end{equation}
in
\begin{equation}
p\,dq - H\,dt = P\,dQ - K\,dt + d(p\,q) - dS
\end{equation}
and comparing coefficients of the differentials, we obtain
\begin{equation}
K(P,\,Q,\,t) = \frac{P^2}{2} - \alpha\,\cos{Q} - \beta\,\cos{(Q-t)}.
\end{equation}
Using the standard trick of adding a dimension to remove the explicit
time dependence, and changing notation as
$$ P\rightarrow p_1, \quad Q\rightarrow q_1, \quad
   K\rightarrow -p_2, \quad t\rightarrow q_2
$$
gives us the Hamiltonian Eq.~(\ref{drivenH}).

This system is roughly analogous to the spin-orbit problem if we take
the inner pendulum as corresponding to the orbit and the outer pendulum
to the spin.

\clearpage

\clearpage

\clearpage

\clearpage

\clearpage

\clearpage

\clearpage

\clearpage

\clearpage

\clearpage

\clearpage

\clearpage

\begin{deluxetable}{ccccccc}
\tablecaption{$H_{k}(e)$ to ${\it O}({e}^4)$ for selected solar system
bodies. Note that $H_2(e)\rightarrow1$ as $e\rightarrow0$.\label{tbl1}}
\tablewidth{0pt}
\tablehead{
\colhead{$k$} & \colhead{Resonance} & \colhead{The Moon} & \colhead{Mercury} &
\colhead{Hyperion} & \colhead{Enceladus} & \colhead{Nereid}}
\startdata
-2&$-1:1$& $3.79\times{10}^{-7}$& $7.73\times{10}^{-5}$& 
$9.83\times{10}^{-6}$& $1.71\times{10}^{-11}$& $1.84\times{10}^{-2}$\\
-1&$-1:2$& $3.45\times{10}^{-6}$& $1.87\times{10}^{-4}$& 
$3.98\times{10}^{-5}$& $1.90\times{10}^{-9}$&$1.22\times{10}^{-2}$\\
1&$1:2$ & $-2.74\times{10}^{-2}$&$-1.02\times{10}^{-1}$&
$-6.17\times{10}^{-2}$&$-2.25\times{10}^{-3}$&$-3.52\times{10}^{-1}$\\
2&$1:1$ & $0.992$& $0.895$& 
$0.962$ & $1.000$ & $-0.149$\\
3&$3:2$ & $1.91\times{10}^{-1}$& $6.54\times{10}^{-1}$&
$4.18\times{10}^{-1}$& $1.57\times{10}^{-2}$&	$-6.18\times{10}^{-1}$\\
4&$2:1$ & $2.54\times{10}^{-2}$& $3.26\times{10}^{-1}$&
$1.25\times{10}^{-1}$&$1.72\times{10}^{-4}$&$-1.28$\\
5&$5:2$ & $2.89\times{10}^{-3}$& $1.39\times{10}^{-1}$&
$3.20\times{10}^{-2}$&$1.60\times{10}^{-6}$& 1.90\\
6&$3:1$ & $3.00\times{10}^{-4}$& $5.36\times{10}^{-2}$&
$7.47\times{10}^{-3}$& $1.37\times{10}^{-8}$&3.27\\
\enddata
\end{deluxetable}

\clearpage

\begin{deluxetable}{ccccccc}
\tablecaption{Physical parameters for selected solar system bodies.\label{tbl2}}
\tablewidth{0pt}
\tablehead{
\colhead{Parameter} & \colhead{The Moon} & \colhead{Mercury} & \colhead{Hyperion} & \colhead{Enceladus} & \colhead{Pandora} 
}
\startdata
$e$ 	 & 0.0549 & 0.206  & 0.1236 & 0.0045 & 0.004 \\
$\alpha$ & 0.026  & 0.0187 & 0.89   & 0.336  & 0.89  \\
\enddata
\tablecomments{The parameter values are taken from the following sources:
The Moon -- \citet{yoder}; Mercury -- \citet{rambauxandbois}; Hyperion -- \citet{blacketal}; Enceladus -- \citet{wisdom2004}; Pandora -- \citet{wisdom1987}.}
\end{deluxetable}

\clearpage


\begin{thebibliography}{}

\bibitem[Binney \& Spergel(1984)]{binneyandspergel} Binney,~J. \&
Spergel,~D. 1984, \mnras, 206, 159

\bibitem[Black et al.(1995)]{blacketal} Black,~G.~J., Nicholson,~P.~D.,
\& Thomas,~P.~C. 1995, Icarus, 117, 149

\bibitem[Blitzer(1979)]{blitzer1979} Blitzer,~L. 1979.  Dynamics of
Orbit-Orbit and Spin-Orbit Resonances: Similarities and Differences, in
{\it Natural and Artificial Satellite Motion, Proceedings of the
International Symposium held at University of Texas at Austin, 1977},
ed. Paul E. Nacozy and Sylvio Ferraz-Mello. (Austin: University of Texas
Press) 

\bibitem[Cary(1981)]{cary} Cary,~J.~R. 1981, Phys. Rep., 79, 129

\bibitem[Cayley(1861)]{cayley} Cayley,~A. 1861, \memras, 29, 191

\bibitem[Celletti(1990a)]{celletti-part1} Celletti,~A. 1990a,
J.~of~Appl.~Math.~and~Phys.~(ZAMP), 41, 174

\bibitem[Celletti(1990b)]{celletti-part2} Celletti,~A. 1990b,
J.~of~Appl.~Math.~and~Phys.~(ZAMP), 41, 453

\bibitem[Celletti et al.(1998)]{cellettietal1998} Celletti,~A.,
Della~Penna,~G. \& Froeschl\'{e},~C. 1998, International Journal of
Bifurcation and Chaos, 8, 2471

\bibitem[Chauvineau \& M\'{e}tris(1994)]{chauvineauandmetris1994}
Chauvineau,~B. \& M\'{e}tris,~G. 1994, Icarus, 109, 191

\bibitem[Chirikov(1979)]{chirikov} Chirikov,~B.~V. 1979, \physrep, 52,
263

\bibitem[Colombo(1965)]{colombo} Colombo,~G. 1965, \nat, 208, 575

\bibitem[Correia \& Laskar(2004)]{correiaandlaskarnature}
Correia,~A.~C.~M. \& Laskar,~J. 2004, \nat, 429, 848

\bibitem[Escande(1982)]{escande} Escande,~D.~F. 1982, \physscr, T2, 126

\bibitem[Goldreich \& Peale(1966)]{goldreichandpeale1} Goldreich,~P. \&
Peale,~ S.~J. 1966, \aj, 71, 425

\bibitem[Goldreich \& Peale(1968)]{goldreichandpeale2} Goldreich,~P. \&
Peale,~ S.~J. 1968, \araa, 6, 287

\bibitem[Goldstein(1980)]{goldstein} Goldstein,~H. 1980.
Classical Mechanics, 2nd Edition.
{\it Addison-Wesley Pub. Co.}, Reading, Mass.

\bibitem[Gustavson(1966)]{gustavson} Gustavson,~F.~G. 1966, \aj, 71, 670

\bibitem[Klavetter(1989)]{klavetter} Klavetter,~J.~J. 1989, \aj, 97, 570

\bibitem[Murray \& Dermott(1999)]{murrayanddermott} Murray,~C.~D. \& Dermott,~S.~F. 1999,
Solar System Dynamics, Cambridge University Press, Cambridge

\bibitem[Pettengill \& Dyce(1965)]{pettengillanddyce}
Pettengill,~G.~H. \& Dyce,~R.~B. 1965, \nat, 206, 1240

\bibitem[Rambaux \& Bois(2004)]{rambauxandbois} Rambaux,~N. \&
Bois,~E. 2004, \aap, 413, 381

\bibitem[Sussman \& Wisdom(2001)]{sussmanandwisdom} Sussman,~G.~J. \&
Wisdom,~J. 2001, Structure and Interpretation of Classical Mechanics,
MIT Press, Cambridge, Massachusetts

\bibitem[Wisdom et al.(1984)]{wpm} Wisdom,~J., Peale,~S.~J. \&
Mignard,~F. 1984, Icarus, 58, 137

\bibitem[Wisdom(1987)]{wisdom1987} Wisdom,~J. 1987, \aj, 94, 1350

\bibitem[Wisdom(2004)]{wisdom2004} Wisdom,~J. 2004, \aj, 128, 484

\bibitem[Yoder(1995)]{yoder} Yoder,~C.~F. 1995, Astrometric and Geodetic
Properties of Earth and the Solar System, in {\it Global Earth
Physics. A Handbook of Physical Constants}, ed. T. Ahrens (American
Geophysical Union, Washington) \end{thebibliography}
\end{document}